\documentclass[11pt]{article}
\usepackage{amsmath,epsf,amssymb,latexsym,enumerate,cite,shadow,array,color}
\usepackage{slashed}
\usepackage{graphicx,wasysym}
 \pdfoutput=1

\DeclareFontFamily{U}{rsf}{} \DeclareFontShape{U}{rsf}{m}{n}{
  <5> <6> rsfs5 <7> <8> <9> rsfs7 <10-> rsfs10}{}
\DeclareMathAlphabet\Scr{U}{rsf}{m}{n} \makeatletter
\@addtoreset{equation}{section} \makeatother

\def\be{\begin{equation}}
\def\ee{\end{equation}}
\def\ba{\begin{array}}
\def\ea{\end{array}}

\newcommand{\bea}{\begin{eqnarray}}
\newcommand{\eea}{\end{eqnarray}}
\def\N{$\cal N$}

\def\K{K{\"a}hler}
\def\Re{\mathop{\rm Re}\nolimits}
\def\Im{\mathop{\rm Im}\nolimits}
\newcommand{\hc}{{\rm h.c.}}
\newcommand{\ft}[2]{{\textstyle\frac{#1}{#2}}}
\def\rmi{{\rm i}}
\def\rmd{{\rm d}}
\def\rme{{\rm e}}
\newcommand{\edet}{e\,}

\usepackage{ifpdf}
\ifx\pdfoutput\undefined
   \pdffalse
\else
   \pdfoutput=1
   \pdftrue
  \usepackage[pdftex]{hyperref}
\newcommand{\gamfive}{\gamma_*}
\newcommand{\mk}{\mathcal{K}}
\newcommand{\ml}{\mathcal{L}}
\newcommand{\mr}{\mathcal{R}}

\begin{document}

\begin{titlepage}

\begin{flushright}
CERN-PH-TH/2010-182\\
SU-ITP-2010-15
\end{flushright}

\begin{center}
{\LARGE \textbf{Superconformal Symmetry, NMSSM, and Inflation
\vskip 0.8cm }}

{\bf {\bf Sergio Ferrara$^{1,2}$},
{\bf Renata Kallosh$^3$}, {\bf Andrei Linde$^3$},\\ {\bf Alessio Marrani$^3$}, and  {\bf Antoine Van Proeyen$^4$} \\

\

$^1${\sl Physics Department, Theory Unit, CERN,  CH 1211, Geneva
23,
Switzerland }\\
$^2${\sl INFN - Laboratori Nazionali di Frascati, Via Enrico
Fermi 40, 00044 Frascati, Italy}\\
$^3${\sl Department of Physics, Stanford University, Stanford, California
94305 USA}\\
$^4${\sl Instituut voor Theoretische Fysica, Katholieke Universiteit
Leuven,\\ Celestijnenlaan 200D, B-3001 Leuven, Belgium}}
\end{center}

\begin{abstract}

We identify a particularly simple class of supergravity models describing superconformal coupling of matter to supergravity. In these models, which we call the canonical superconformal supergravity (CSS) models, the kinetic terms in the Jordan frame are canonical,
and the scalar potential is the same as in the global theory.  The pure supergravity part of the total action has a local Poincar{\'e} supersymmetry, whereas the chiral and vector multiplets coupled to supergravity have a larger local superconformal symmetry.

The scale-free globally supersymmetric theories, such as the NMSSM with a scale-invariant superpotential, can be naturally embedded into this class of theories. After the supergravity embedding, the Jordan frame scalar potential of such theories remains scale free; it is quartic, it contains no mass terms, no nonrenormalizable terms, no cosmological constant.

The local superconformal symmetry can be broken by additional terms,
which, in the small field limit, are suppressed by the gravitational
coupling. This can be achieved by introducing the nonminimal
scalar-curvature coupling, and by taking into account interactions with a
hidden sector.  In this approach,  the smallness of the mass parameters
in the NMSSM may be traced back to the original superconformal
invariance. This allows one to address the $\mu$ problem and the cosmological
domain wall problem in this model, and to implement chaotic inflation in
the NMSSM. We discuss the gravitino problem in the NMSSM inflation, as
well as the possibility to obtain a broad class of new versions of
chaotic inflation in supergravity.
\end{abstract}

\vspace{24pt}
\end{titlepage}

{\parskip -6pt \tableofcontents}

\parskip 9pt

\section{Introduction}

This work is a continuation of our previous paper \cite{Ferrara:2010yw}
where we studied generic supergravity in the Jordan frame and the
possibility to implement the Higgs-type inflation  \cite{Sha-1} in the
context of the next-to-minimal supersymmetric standard model (NMSSM)
\cite{Einhorn:2009bh,Ferrara:2010yw,Lee:2010hj}.

These recent developments were based on a combination of long efforts of many authors in several seemingly unrelated directions.

1) Many decades ago, one of the most popular ways to describe
gravitational interactions of a  scalar field $\varphi$ was to assume that it is
conformally coupled to gravity, which means that its Lagrangian contains a term $-\sqrt{-{g}}\, { \varphi^2\over 12}
R$; see e.g. \cite{BD}. With the invention
of inflation, which was difficult to achieve for conformally coupled
scalars, the concept of scalar fields conformally coupled to gravity
gradually lost part of its appeal. On the other hand, several authors
emphasized that inflation may occur in a very natural way if a scalar
field nonminimally couples to gravity, with a sign opposite to that of the conformal coupling; see e.g.
\cite{Salopek:1988qh}. Recently there was a revival of interest in this
possibility after it was realized that it may allow inflation in the
standard model, with the Higgs field playing the role of the inflaton
\cite{Sha-1}. However, for a while it was not clear whether one could
implement this idea in supersymmetric generalizations of the standard
model. Some progress in this direction was reached only very recently
\cite{Einhorn:2009bh,Ferrara:2010yw,Lee:2010hj}. In this paper we will
develop a more systematic approach to this issue.

2) Conformal invariance plays an important role in the formulation of supergravity.  The general formulation of supergravity starts with the superconformal theory. Then, after gauge fixing, which, in particular, makes the conformal compensator field proportional  to the Planck mass, one derives the standard textbook formulation of supergravity
\cite{Cremmer:1978hn,BFNS-82,CFGVP-1,Girardi:1984eq,Wess:1992cp,Weinberg:2000cr,Kallosh:2000ve}. Once this step is made, the theory is formulated in the Einstein frame, all scalars have minimal coupling to gravity, and the superconformal origin of supergravity becomes well hidden.

In \cite{Ferrara:2010yw} we performed an alternative gauge fixing of the version of the superconformal theory,
which allows one to derive the supergravity action in an arbitrary Jordan frame. This provides a complete locally supersymmetric theory for scalars with a  nonminimal coupling to gravity.

3) Prior to the discovery of supergravity, the development of particle physics was successfully guided by the principle of gauge invariance and renormalizability. However, supergravity is nonrenormalizable. In general, one can write any kind of superpotential which may lead to nonrenormalizable interactions which become important even at low energy. It would be nice to have a formulation of supergravity where the low-energy renormalizability appears as a result of some general principle, similar to the principle of spontaneously broken gauge invariance.

The extraordinary smallness of the Higgs mass, as compared to the Planck
mass, can be protected by supersymmetry, but only if the Higgs mass was
extremely small to start with. The minimal supersymmetric standard model
(MSSM) and the general NMSSM include several other dimensional parameters
which are required to be extraordinary small ($\mu$-problem, tadpole
problem). These issues can be addressed in the context of the
$\Bbb{Z}_{3}$-invariant NMSSM, which requires that the superpotential
describing the standard model is scale invariant \cite{Ellwanger:2009dp}.
However, it would be important to find some fundamental underpinnings of
this requirement. Moreover, $\Bbb{Z}_{3}$ symmetry of the scale-invariant
superpotential leads to the cosmological domain wall problem.

All of these problems have been discussed extensively in the existing
literature, but recent developments stimulated us to look at these
issues again in \cite{Ferrara:2010yw}, returning back to the
superconformal origin of supergravity. As we will see, many of these
problems become much easier to address in a class of models where the
original superconformal invariance remains at least partially preserved,
being broken only by gravitational effects, or by  anomalies.
 This symmetry may naturally explain renormalizability and smallness of the mass parameters in the standard model. It leads to a formulation of supergravity in the Jordan frame, where in certain cases
the potentials and kinetic terms look as simple as in the global supersymmetry (SUSY). In this context, one can achieve inflation and simultaneously solve the domain wall problem with the help of the terms describing scalars nonminimally coupled to gravity.

To explain our main idea, let us consider a nonsupersymmetric conformally invariant toy model describing gravity and two real scalar fields, $\phi$ and $h$:
\begin{equation}
\mathcal{L} = \sqrt{-{g}}\left[{1\over 2}\partial_{\mu}\phi \partial_{\nu}\phi \, g^{\mu\nu}  +{ \phi^2\over 12}  R({g})- {1\over 2}\partial_{\mu} h \partial_{\nu} h \, g^{\mu\nu}  -{h^2\over 12}  R({g}) -{\lambda\over 4}h^4\right]\,.
\label{toy}
\end{equation}
The field $\phi(x)$ is referred to as  a conformal compensator. This
theory is locally conformal invariant under the following
transformations: \be g'_{\mu\nu} = \rme^{-2\sigma(x)} g_{\mu\nu}\,
,\qquad \phi' =  \rme^{\sigma(x)} \phi\, ,\qquad h' =  \rme^{\sigma(x)}
h\ . \label{conf}\ee Note that the kinetic term of the conformal
compensator $\phi$ has a wrong sign. This is not a problem because there
are no physical degrees of freedom associated with it; the field $\phi$
can be removed from the theory by fixing the gauge symmetry
(\ref{conf}). If we choose the gauge $\phi(x) = \sqrt{6}\,  M_P$, the
$\phi$-terms in (\ref{toy}) reduce to the Einstein action. The full
Lagrangian in the Jordan frame is
\begin{equation}
\mathcal{L}_{\rm total }= \mathcal{L}_{\rm E } + \mathcal{L}_{\rm conf }= \sqrt{-{g}}\,{M_P^2\over 2}  R({g})-  \sqrt{-{g}}\left[{1\over 2}\partial_{\mu} h \partial_{\nu} h \, g^{\mu\nu}  +{ h^2\over 12} R({g}) +{\lambda\over 4}h^4\right]\,.
\label{toy2}
\end{equation}
It consists of two parts, the Einstein Lagrangian $
\sqrt{-{g}}\,{M_P^2\over 2}  R({g})$, which is not conformally
invariant, and the conformally invariant theory of the canonically
normalized scalar field $h$,
\be
 \mathcal{L}_{\rm conf }=-  \sqrt{-{g}}\left[{1\over 2}\partial_{\mu} h \partial_{\nu} h \, g^{\mu\nu}  +{ h^2\over 12} R({g}) +{\lambda\over
 4}h^4\right]\,.
\label{c}
\ee

As we already mentioned, theories of this type played a very important
role in the development of particle physics and cosmology many decades
ago; see e.g. \cite{BD}.  One of the main reasons is that the Friedmann
universe is conformally flat. By making a conformal transformation, one
could represent equations of motion of the scalar field in the Friedmann
universe in terms of equations of motion of a conformally transformed
field in Minkowski space, which is a tremendous simplification.

The theory (\ref{toy}) is unique if we require that the local conformal
symmetry of the $h$ part of the action, which has canonical kinetic
terms, should be preserved after the gauge fixing. It is determined by
the condition that the conformal compensator $\phi(x)$ is decoupled from
the field $h(x)$.

The conformal symmetry of the matter action in Eq. (\ref{c}) is manifest in the Jordan frame (\ref{toy2}).
One can make a certain field and metric transformation and switch to the Einstein frame, where the term $-  \sqrt{-{g}}{ h^2\over 12} R({g})$ is absorbed into the Einstein action. This allows one to use the standard Einstein equations. However, after this transformation both the gravity part as well as the matter part of the action have conformal symmetry broken.

Similarly, the standard formulation of supergravity interacting with matter brings us directly to the Einstein frame, where the original superconformal symmetry is lost even in the special class of models where matter fields are decoupled from the conformal compensator. That is why it was hard to see any advantages of this class of  models in the standard textbook formulation of supergravity.  Meanwhile, as we will see shortly, in the class of models with conformal coupling of scalars, the matter Lagrangian in the Jordan frame looks exceptionally simple: all kinetic terms are canonical in the simplest case, the superpotential contains only cubic terms, and the scalar potential is quartic with respect to the scalar fields, just as in our toy model (\ref{toy2}). The theories of this class provide a very natural supergravity embedding of the $\Bbb{Z}_{3}$-invariant NMSSM with a scale-free superpotential.

Of course, at the end of the day we want to make most of the particles
massive. Thus, we would need to break superconformal invariance, but we
would like to do it in a way that preserves some of the most attractive
features of the original superconformal theory.

Superconformal symmetry may be broken by anomalies, by interaction with
a hidden sector, or by gravitational interactions suppressed by inverse
powers of the Planck mass. However, one should try to avoid introducing
into the original theory any terms proportional to the compensator
field. For example, one could add to (\ref{toy}) a term
${c^2}h^{2}\phi^2$ without breaking the conformal invariance of the
model (\ref{toy}). However, in the gauge $\phi = \sqrt{6} M_P$, the
matter part of the theory (\ref{toy2}) would acquire the term $6 \, c^2
M_P^2h^2$. This term strongly breaks the conformal invariance by giving
the field $h$ a mass squared $12 \, c^2 M_P^2$, which is enormously
large unless the dimensionless constant $c$ is extremely small. This
would lead to the hierarchy problem if one tries to use the field $h$
for the description of the low-energy physics. Similarly, the term $c\,
\phi^3 h$ would introduce a huge tadpole $\sim M_P^3 h$, and the term
$\phi^4$ would introduce an enormously large cosmological constant $\sim
M_P^4$. We would like to use the original (super)conformal symmetry to
protect us against such problems.

On the other hand, the terms $h^{4+n}/\phi^n$, which are inversely
proportional to $\phi^n$, would lead to nonrenormalizable interaction
terms $\sim h^{4+n}/M_P^n$. While such terms are unpleasant, they
usually appear in the Einstein frame anyway, and they are not expected
to affect particle physics at energies and fields much smaller than
$M_P$.

In this paper, we will break the local superconformal symmetry of the
matter coupling via the real part of the quadratic holomorphic function
in scalar-curvature coupling, by the terms in the \K\, potential which
are suppressed by the inverse Planck mass, and also by the interaction to
the hidden sector. This produces additional terms in the action. At small
fields and low energies, the new terms are suppressed by inverse powers
of the Planck mass. In other words, at small fields and low energies, the
original superconformal symmetry is broken only by  effects suppressed by
the small gravitational coupling. This can be very helpful in particle
phenomenology. The smallness of the new terms helps to explain the
smallness of the Higgs mass and of the $\mu$ term, which appear only
because of the breaking of the superconformal invariance. On the other
hand, these terms can be large enough to address the domain wall problem
in the NMSSM. Moreover, in the large field limit, some of the new terms
become dominant and allow one to implement inflation in supergravity
along the lines of  \cite{Sha-1,Salopek:1988qh} and
\cite{Einhorn:2009bh,Ferrara:2010yw,Lee:2010hj}.

The paper is organized as follows. In Sec. \ref{old} we give a short
summary of the results of Ref. \cite{Ferrara:2010yw} on the supergravity
action in an arbitrary Jordan frame defined by the function of scalars
$\Phi(z, \bar z)$. We then focus on a special case of $ \Phi \left(
z,\bar z\right) =
-3M_{P}^{2}\rme^{-[\mathcal{K}\left( z,\bar z \right)/3M_{P}^{2}]}=
-3M_{P}^{2}+\delta _{\alpha \bar \beta }z^{\alpha }\bar
z^{\bar \beta }+J\left( z\right) +\bar J\left( \bar z\right)$, where
$J(z)$ is a holomorphic function quadratic in $z$. It has been found in
\cite{Ferrara:2010yw} that in this case the kinetic term for scalars is
canonical since it is defined by $ \Phi _{\alpha\bar\beta }= \delta
_{\alpha\bar\beta }$. We explain the role of the auxiliary supergravity
vector fields $ A_\mu$.

Section \ref{scmatter} presents the simplest possible embedding of the
globally superconformal theory into supergravity. We start with the
analog of (\ref{toy}), the theory with the  $SU(2,2|1)$ local
superconformal symmetry and no dimensional parameters. It contains an
extra chiral multiplet, the compensator one. The theory is based on the
results obtained in \cite{CFGVP-1,BFNS-82}, and more recent results of
Refs. \cite{Kallosh:2000ve} and \cite{Ferrara:2010yw}.  We then specify
the superconformal action for the case when (a) all kinetic terms are
canonical and (b) the matter multiplets decouple from the compensator. We
find that in these models the total supergravity action consists of the
pure supergravity part, which breaks superconformal symmetry, and the
matter part, which remains superconformal after the gauge fixing. The
scalars are conformally coupled to gravity, kinetic terms are canonical,
and the supergravity potential coincides with the global theory
potential.  We call these theories {\it canonical superconformal
supergravity} (CSS) models.

In this sense, the embedding of the globally superconformal theory into supergravity in the Jordan frame  becomes a simple additive operation: One adds the action of the global SUSY, interacting with gravity with conformal scalar-curvature coupling, to the action of supergravity; that is it. However, this simple operation looks much more complicated in the Einstein frame.

We further develop a geometric way to break the superconformal coupling
of matter to supergravity. The flat \K\, geometry of the chiral
multiplets, including the compensator field, is replaced by a nonflat
geometry without introducing any new dimensional parameters.
Specifically, we study CSS models with superconformal symmetry broken by
the $\chi$ term: the real part of the holomorphic function defining the
Jordan frame. This leads to useful applications both in particle physics
and cosmology.

In  Sec. \ref{ss:sugraNMSSM} we apply the method of embedding globally
supersymmetric theories into supergravity described in Sec. \ref{scmatter} to the
scale-invariant version of the NMSSM. Section \ref{phen} has a short discussion of
the issues of the NMSSM phenomenology, including the $\mu$-problem and
domain wall problem. We argue that using the superconformal matter action
with the $\chi$ term in combination with a hidden sector may resolve both
of these problems.

Sections \ref{ss:Higgstypenonmin} and \ref{ss:InflsugraNMSSM} are devoted
to inflation. We first review the Higgs-type inflation in the standard
model, following \cite{Sha-1}. We argue that this inflationary model, as
well as its NMSSM generalization proposed in \cite{Einhorn:2009bh} and
developed in \cite{Ferrara:2010yw,Lee:2010hj}, does not suffer from the
problems related to the unitarity bound discussed in
\cite{Burgess:2009ea,Barbon:2009ya,Hertzberg:2010dc,Germani:2010gm}. We
describe observational implications of inflation in the NMSSM and find
that these implications are invariant with respect to a certain rescaling
of the parameters of this model. We describe a  mechanism of
stabilization of the inflationary trajectory, which is necessary for
consistency of this scenario. It includes a requirement of special
corrections which stabilize some moduli at the origin of the moduli
space. We discuss the gravitino problem, which may appear in this
scenario, and point out the existence of a broad class of new
inflationary models based on the ideas described in our paper, where the
inflaton is not necessarily related to the Higgs field and the gravitino
problem may not arise \cite{Kallosh:2010ug}.

Our results are briefly summarized in the Conclusions.
Appendix~\ref{app:completeAct} presents a complete action of
superconformal matter coupled to gravity, including vector multiplets and
fermions with canonical kinetic terms.  It is given in Eqs.
(\ref{SSGwithtorsion})-(\ref{valueFD}) which provide the generalization
of the scalar-gravity action in Eqs. (\ref{L0}), (\ref{VJ0}), when all
fermions and vectors are included. Appendix~\ref{app:Why}  shows how one
can derive the simple CSS potential, which is the same as in global SUSY
theories, starting from the generic Einstein frame supergravity
potential. Appendix~\ref{app:G} provides the metric of the moduli space
for CSS models with superconformal symmetry broken by the real part of
the holomorphic function. Appendix~\ref{app:hands} presents a detailed
expression for the potential of the  scalar field $s$ and the inflaton
field $h$ in the Jordan frame and in the Einstein frame.

Thus, the paper essentially consists of two parts. Those who are
interested mainly in phenomenological and cosmological implications of
our construction, may take a quick look at Secs \ref{old},
 \ref{scmatter},  \ref{ss:sugraNMSSM}, and \ref{ss:sugraNMSSM} and then proceed directly to
Secs. \ref{phen},  \ref{ss:Higgstypenonmin}, and \ref{ss:InflsugraNMSSM}.
However, we believe that the superconformal approach to supergravity and
the CSS models described in Secs.  \ref{old}, \ref{scmatter}, and
\ref{ss:sugraNMSSM} deserve further investigation quite independently of
their immediate implications for inflation and the NMSSM.

\section{Supergravity in the Jordan frame}
\label{old}

 The general theory of supergravity in an arbitrary Jordan frame was   derived in \cite{Ferrara:2010yw} by a gauge fixing of the $SU(2,2|1)$ superconformal theory \cite{Kallosh:2000ve}.  This approach  is based on earlier work on superconformal origin of the supergravity theory in \cite{Kaku:1977pa}.
The extra gauge symmetries of the superconformal theory, including a
local conformal symmetry, which rescales the metric, allow
a possibility to derive the supergravity action either in the Einstein
frame or  in an arbitrary Jordan frame.  The Einstein frame Lagrangian,
in units of $M_P=1$,   is ${\cal L}_E= \sqrt{-g_E}\  { 1\over 2} \,R
(g_E)+...$, there is no direct scalar-curvature coupling. The Jordan
frame Lagrangian is ${\cal L}_J= -\sqrt{-g_J}\,  { \Phi(z, \bar z)\over
6} \, R (g_J)+...$, where $\Phi(z, \bar z)$ is an arbitrary function of
complex scalar fields  $z, \bar z$. Therefore, in general, there is a
scalar-curvature coupling in the Jordan frame. A local conformal
symmetry allows to make a choice of $\Phi(z, \bar z)=-3$ to get the
Einstein frame supergravity. Otherwise with the  frame function
depending on scalars we get the Jordan frame supergravity. The relation
between the space-time metrics is given by $g^{\mu\nu  } _E= \Omega^2(z,
\bar z)  g^{\mu\nu }_J$,  where $\Omega^2(z, \bar z)=-{1\over 3}
\,\Phi(z, \bar z)$.

The  scalar-gravity part of the ${\cal N}=1$, $d=4$ supergravity in a generic Jordan frame with frame function $\Phi(z, \bar z)$, a \K\, potential $\mathcal{K}(z, \bar z)$ independent on the frame function, and superpotential $W(z)$ is, according to \cite{Ferrara:2010yw},
\begin{eqnarray}
\mathcal{L}_{J}^{\rm scalar-grav} = \sqrt{-{g}_{J}}\left[\Phi \left(  -\frac{1}{6} {R}({g}_J)+ {\cal A}^2_\mu(z, \bar z) \right )
+\left( \frac{1}{3}\Phi g_{\alpha\bar\beta }-\frac{\Phi
_\alpha \Phi _{\bar\beta }}{\Phi }\right) \hat {\partial}_\mu z^\alpha \hat {\partial}^\mu \bar z^{\bar\beta }
-V_J \right]\,.
  \label{Turin-4}
\end{eqnarray}
Here
\begin{eqnarray}
&& \Phi_\alpha \equiv {\frac{\partial }{\partial z^\alpha }}\Phi(z, \bar z) \,, \qquad  \Phi_{\bar\beta } \equiv {\partial\over \partial{\bar z}^{\bar\beta }}  \Phi(z, \bar z)=\overline{\Phi} _{\bar \beta} \, ,
\nonumber\\
&& g_{\alpha \bar \beta}= {\frac{\partial^2  \mathcal{K}(z, \bar z)}{\partial z^\alpha  \partial \bar z^{\bar \beta}}}\equiv \mathcal{K}_{\alpha \bar \beta} (z, \bar z)\, ,
\end{eqnarray}
and $\mathcal{A}_\mu$ is the purely bosonic part of the on-shell value of the auxiliary field $A_\mu $. On shell it depends on scalar
fields as follows:
\be {\cal A}_\mu(z,{\bar z}) \equiv
-\frac{\rmi}{2\Phi }\,
  \left( \hat {\partial}_\mu z^\alpha\partial_\alpha\Phi
  - \hat {\partial} _\mu \bar z^{\bar\alpha }\partial_{\bar\alpha }\Phi \right).
 \label{A} \ee
The gauge covariant derivative  $\hat {\partial}_\mu z^\alpha $ in Eqs. (\ref{Turin-4}), (\ref{A}) is
\begin{eqnarray}
\hat {\partial}_\mu z^\alpha & \equiv  & \partial_\mu z^\alpha  - A_\mu^A k_A^\alpha \, , \label{covderconf}
\end{eqnarray}
where $A_\mu^A$ is the vector gauge field and  $k_A^\alpha $ is the Killing vector, defining the gauge transformations of  scalars, $\delta z^\alpha= \theta^A  k_A^\alpha$.
The Jordan frame potential \begin{equation}
V_{J}=\frac{\Phi ^{2}}{9}V_{E}\,  \label{Turin-3}
\end{equation}
is defined via the Einstein frame potential
\begin{equation}
V_{E}=V_{E}^F +V_{E}^D = \rme^{\mathcal{K}}\left( -3W\overline{W}
+\nabla _\alpha Wg^{\alpha \bar\beta }\nabla _{\bar\beta }\overline{W}
\right) +  \ft12(\Re
f)^{-1\,AB}P_AP_B\,, \label{VE}
\end{equation}
where $\nabla _\alpha W$ denotes the K{\"a}hler-covariant
derivative of the superpotential and $P_A$ is a momentum map.
A special important class of the superconformal models with
\begin{equation}
 \Phi(z,{\bar z}) = -3\rme^{-\ft13 \mathcal{K}(z,{\bar z})} \label{Turin}
\end{equation}
and the  corresponding actions in the Jordan frame
were derived in components in \cite{Cremmer:1978hn,CFGVP-1}, and in
superspace in \cite{Girardi:1984eq,Wess:1992cp}.
In this case the  simpler form of $\mathcal{L}
_{J} $ given by (\ref{Turin-4}) was found in \cite{Ferrara:2010yw}:
\begin{equation}
\mathcal{L}_{J}=\sqrt{-g_{J}}\left[ \Phi\left( -
\frac{1}{6} R(g_{J})+
{\cal A}^2_\mu(z, \bar z)  \right)
-\Phi _{\alpha\bar\beta }\hat{\partial}
_\mu z^\alpha\hat{\partial}^\mu \bar z^{\bar\beta }-\frac{\Phi^2}{9}V_{E}\right] \, .  \label{action}
\end{equation}
Also an interesting observation about the Jordan frame kinetic terms for scalars was made:
{\it For a particular choice of the frame function the kinetic scalar terms are canonical} when the on-shell auxiliary axial-vector field ${\cal A}_\mu$ vanishes. This requires that
\begin{equation}
 \Phi(z,{\bar z}) = -3\rme^{-(1/3) \mathcal{K}(z,{\bar z})} = -3 + \delta_{\alpha\bar\beta }z^{\alpha} \bar z^{\bar\beta } +J(z) +\bar J({\bar z})\,  \ ,
\label{Phi}
\end{equation}
and it follows that
\be
 \Phi _{\alpha\bar\beta }\equiv {\frac{\partial^2  \Phi(z, \bar z)}{\partial z^\alpha  \partial \bar z^{\bar \beta}}}= \delta _{\alpha\bar\beta } \ ,
\label{Phi''}
\end{equation}
where $\mathcal{K}(z,{\bar z})$ is the \K \, potential and $J(z)$ is holomorphic.
For the choice (\ref{Phi}) the action in the Jordan frame is
\begin{equation}
{\mathcal{L}_{J}\over \sqrt{-g_J}\,}= \Phi (z,{\bar z})\left(-{\textstyle
\frac{1}{6}} R (g_J) +{\cal A}^2_\mu(z, \bar z) \right)-
\delta _{\alpha\bar\beta }\hat {\partial}_\mu z^\alpha \hat {\partial}^\mu \bar z^{\bar\beta }
- V_J(z,{\bar z})\,,  \label{kin}
\end{equation}
where ${\cal A}_\mu(z,{\bar z})$ is defined in Eq. (\ref{A}). It
 vanishes in many cosmological applications with either real or imaginary scalar fields. For such configurations with ${\cal A}_\mu=0$ the second  term in Eq. (\ref{kin}) is a canonical kinetic term for scalars. This simplification of the supergravity theory in the Jordan frame with regard to kinetic terms of scalars is, as we will see below, a particular property of the class of supergravity theories which have a  superconformal  matter-supergravity coupling.

\section{Superconformal  matter coupling  in the Jordan frame supergravity}\label{scmatter}

\subsection{Locally  superconformal theory}

Superconformal theory is the starting point to derive supergravity.  We
will consider here a class of models where the chiral and vector multiplets
do not interact with the superconformal compensator field. This will
provide a simple embedding of globally supersymmetric models into
supergravity in the Jordan frame. We will further introduce a geometric
mechanism of breaking of the superconformal symmetry, which is suitable
for phenomenology and cosmology. The superconformal symmetry will also
be broken by radiative corrections and by terms suppressed by inverse
powers of $M_P^2$.

To embed  a given globally supersymmetric model  into supergravity, a  particular \K\, potential has to be chosen, which could be any real function of all scalars (with positive definite metric of moduli space). In the Einstein frame, where there is no direct coupling of curvature to scalars,  the kinetic term for scalars is $\ml_{\rm kin} =\mk_{\alpha \bar \beta} (z, \bar z)\partial z^\alpha \partial \bar z^{\bar \beta}$ and the $F$-term potential in the Einstein frame is  $V_E^F={\rm e}^{\mathcal{K}}\left(
\nabla _\alpha W\mk ^{\alpha \bar\beta }\nabla _{\bar\beta }\overline{W} -3W\overline{W}
\right)$
where $\nabla _\alpha W$ denotes the K{\"a}hler-covariant
derivative of the superpotential.  If one would like to preserve the property of the globally supersymmetric theory to have canonical kinetic terms, one has to take  $\mk = \delta_{\alpha \bar \beta} z^\alpha \bar z^{\bar \beta}$,  up to \K\, transformation. But in such case the $F$-term potential is quite different from the global supersymmetry case  $V_{\rm global}=|\partial W |^2$.

Here we would like to present a  simple  case of the embedding of a
class of scale-invariant globally supersymmetric models into
supergravity. Embedding, in general, means that the total action of
supergravity and chiral and vector \N=1 multiplets has a local Poincar{\'e}
supersymmetry.  A special class that we will present here has the
property in which the part of the action describing chiral and vector
multiplets coupled to supergravity has a much larger local
superconformal symmetry.  This symmetry is broken down to the local
Poincar{\'e} supersymmetry only by the part of the action describing the
self-interacting supergravity multiplet.  First we start with the analog
of (\ref{toy}), the theory with the  $SU(2,2|1)$ superconformal symmetry
and no dimensional parameters: it contains an extra chiral multiplet,
the compensator one. The theory is based on Refs.
\cite{CFGVP-1,BFNS-82} and \cite{Kallosh:2000ve,Ferrara:2010yw}.  We
then specify   the superconformal action for the case when the matter
multiplets decouple from the compensator. As a result, we find in this
class of theories that the total supergravity action consists of the
pure supergravity part, which breaks superconformal symmetry, and the
matter part which remains superconformal after the gauge fixing.

The supergravity Weyl multiplet consists of the vierbein, gravitino and the vector gauge field of the $U(1)\, \mr$ symmetry:
 $ e_\mu ^a, \psi_\mu$, and  $A_\mu$. The chiral multiplet has scalars and spinors, and the vector multiplet has gauge fields and gauginos.

We start here with the superconformal action described in detail   in
\cite{Kallosh:2000ve} in Eqs. (3.3)-(3.8) and more recently \footnote{The
action in Eqs. (3.3)-(3.8) of \cite{Kallosh:2000ve} has an auxiliary
vector field $A_\mu$ as an independent field, before the equations of
motion have been used. In Sec. 5.1 of \cite{Ferrara:2010yw} the
superconformal action has $A_{\mu }$ already on shell, with purely
bosonic and fermionic parts, respectively given by the first and second
expressions of Eq. (5.13) of \cite{Ferrara:2010yw}. Here it is important
for us to keep $A_\mu$ as an off-shell independent field.} in
\cite{Ferrara:2010yw}  in Sec. 5.1. This action has a local $SU(2,2|1)$
superconformal symmetry and no dimensional parameters. The symmetries
include local dilatation, special conformal symmetry, special
supersymmetry, and local $U(1)\, \mr$ symmetry, in addition to all local
symmetries of supergravity. The special conformal symmetry has an
independent field $b_\mu$ as a gauge field, and the  local $U(1)\, \mr$
symmetry has $A_\mu$ as a gauge field. When the local  dilatation,
special conformal symmetry, special supersymmetry and local $U(1)\, \mr$
symmetry  are gauge fixed, one finds a generic supergravity theory
without extra symmetries. In notation of \cite{Ferrara:2010yw} the action
contains 3 superconformal-invariant terms
\begin{equation}
\mathcal{L}_{\rm sc}= [\mathcal{N}(X,\bar X) ]_D + [\mathcal{W}(X)]_F + \left[f_{AB }\bar\lambda^A P_L \lambda^B\right]_F  \label{sconfaction}
\,.
\end{equation}
The first one codifies the K{\"a}hler potential
for the $n+1$ superconformal fields $X^I, \bar X^{\bar J}$, the second introduces a superpotential, and the third involves the chiral kinetic matrix
$f_{AB}(X)$ (where $A,B$ are the gauge indices), and gauginos $\lambda^A$,
 $P_L$ projects on the left-handed
fermions. The superconformal chiral multiplets contain the bosonic fields $X^I$,
 fermions $\Omega^I$,  and auxiliary fields $ F^I$, $I=0, 1,..., n$. One of the multiplets, $X^0, \Omega^0, F^0$, can be viewed as a compensator multiplet. Its purpose is to provide the part of the local superconformal symmetries which are absent in supergravity.

 The dilatation symmetry implies $\mathcal{N}\left( X, \bar X\right)$ to be homogeneous of first degree in both $X$ and $\bar X$, $\mathcal{W}\left( X\right) $
to be homogeneous of third degree in $X$, and $f_{AB}\left(X\right)$ to
be homogeneous of zeroth degree in $X$. Under chiral $U(1)\, \mr$
symmetry $\mathcal{N}(X, \bar X)$ and $f_{AB}(X)$ are neutral,
$\mathcal{W}(X)$ has chiral weight 3, and $\overline  { \mathcal{W}}(\bar
X)$ has  chiral weight -3.

 The $n+1$ scalars including the compensator multiplet form a K{\"a}hler manifold with
metric, connection, and curvature given, respectively,  by
\be
  G_{I\bar J}=\partial_I \partial_{\bar J} {\cal N} \equiv {\partial  \mathcal{N}(X, \bar X)\over \partial X^I \partial \bar X^{\bar J}}
\label{LargeK}
\ee
  and
    $ \Gamma^I_{JK}= G^{I\bar L}{\cal N}_{JK\bar L}
 $, $
  R_{I\bar K J\bar L}= {\cal N}_{IJ\bar K\bar L}-{\cal N}_{IJ\bar M}G^{M\bar M}{\cal N}_{M\bar K\bar L}$.
  For example, the complete
 gravity-scalar part of the $SU(2,2|1)$ invariant superconformal action has a gravity part, kinetic terms for scalars and a potential:
\be
{1\over \sqrt{-g}}\mathcal{L}_{\rm sc}^{\rm scalar-grav}=-\ft16{\cal N} (X,\bar X)R
-G_{I\bar J}D^\mu X^I\,D_\mu \bar X^{\bar J}-V_{\rm sc} \, ,
 \label{scGrav}
\end{equation}
where
\begin{equation}
  V_{\rm sc}\,=\, V_F+ V_D= G^{I\bar J}{\cal W}_I \overline{{\cal W}}_{\bar J} + \ft12(\Re f)^{-1\,AB}  \mathcal{P}_A  \mathcal{P}_B
\,.
 \label{confactpot}
\end{equation}
Here
\be
{\cal W}_I\equiv {\partial {\cal W}\over \partial X^I}\ , \qquad  \overline{{\cal W}}_{\bar J}\equiv  {\partial \overline{{\cal W}}\over \partial \bar X^{\bar J}} \ .
\ee
The $F$-term potential originates from the solution for the auxiliary field for the chiral multiplet,
$F^I= G^{I\bar J} \overline {\cal W}_{\bar J}$. The $D$-term potential originates from the solution for the auxiliary field for the vector multiplet, $D^A=(\Re f)^{-1\,AB} \mathcal{P}_B$, where $\mathcal{P}_A$ is the momentum map defining the $D$-term potential.
The covariant derivative  $D_\mu $ in Eq. (\ref{scGrav}) is
\begin{eqnarray}
D_\mu X^I & = & \partial_\mu X^I-b_\mu X^I -\rmi A_\mu
X^I  - A_\mu^A k_A^I \, , \label{covderconf1}
\end{eqnarray}
where $A_\mu^A$ is the vector gauge field and  $k_A^I$ is the Killing vector, defining the gauge transformations of  scalars $\delta X^I= \theta^A  k_A^I$. We have included the gauge
vectors  $A_\mu^A$ into the covariant derivatives (\ref{covderconf1}) to make clear the relation between gauge symmetries and $D$-term potential where
\be
\mathcal{P}_A= \rmi{\cal N}_I k_A^I= \mathcal{P}_A^{\dagger}\,.
\ee

 \subsection{CSS models }\label{scmodes}

Here we introduce a set of canonical superconformal supergravity
models, starting from the $SU(2,2|1)$ superconformal action. In these
models, after the gauge fixing  of local dilatation, special conformal
symmetry, special supersymmetry, and local $U(1)\, \mr$ symmetry, the
resulting {\it action} for the $n$ chiral multiplets and all vector multiplets
{\it remains superconformal}. The reason for this is that the original
superconformal theory before gauge fixing has all $n$ chiral  multiplets
and vectors multiplets {\it decoupled from the compensator} multiplet
$X^0, \Omega^0$, and $F^0$. We focus here on the {\it simplest version} of such
superconformal matter coupling models when {\it kinetic terms are canonical}.
We therefore define the CSS class of models by the following conditions:

\noindent (1)  We choose\footnote{After the compensator field is gauge-fixed, the manifold of physical $n$ complex scalars becomes that of ${SU(1,n)\over U(n)}$ non-compact space.}
a {\it flat $SU(1,n)$ K{\"a}hler manifold} for all $n+1$ chiral multiplets $X^I$, including the compensator field $X^0$
\be
\mathcal{N}(X,\bar X)= -|X^0|^2 + |X^\alpha|^2 \, , \qquad \alpha =1,...,n\,.
\label{calNminimal}
 \ee
 This means that $\eta_{0\bar 0}=-1,  \, \eta_{\alpha  \bar \beta  }=\delta _{\alpha  \bar \beta }$, where
   \begin{equation}
  G_{I\bar J}={\cal N}_{I\bar J}= \eta_{I\bar J} \,,\qquad G^{I\bar J}= \eta^{I\bar J}\, , \qquad
  \Gamma^I_{JK}= 0\, , \qquad
  R_{I\bar K J\bar L}= 0\,.
 \label{flat}
 \ee
(2) We choose a cubic, {\it $X^0$ independent superpotential}, which breaks the  $SU(1,n)$ symmetry:
 \be
{\cal W}(X)
=  {1\over 3} d_{\alpha \beta \gamma} X^\alpha  X^\beta  X^\gamma \qquad \Rightarrow \qquad {\cal W}_0\equiv {\partial {\cal W} \over \partial X^0}=0 \ .
\ee
(3)  We choose a {\it constant complex vector kinetic matrix} and
$\Re f_{AB} $ is a constant positive definite  matrix.

\noindent (4) We choose an {\it $X^0$ independent momentum map}. The
conformal requirements respecting (\ref{calNminimal}) imply that the transformations are of the
form
\begin{equation}
  k_A^\alpha = (m_A)^\alpha {}_\beta X^\beta \,,\qquad k_A^{\bar \alpha }=(\overline{m}_A)^{\bar \alpha }{}_{\bar \beta }X^{\bar \beta }\,,\qquad
  {\cal P}_A = \rmi\delta _{\alpha \bar \beta }X^{\bar \beta }(m_A)^\alpha {}_\gamma  X^\gamma =
  -\rmi\delta _{\alpha \bar \beta }X^\alpha (\overline{m}_A)^{\bar \beta }{}_{\bar \gamma }\bar X^{\bar \gamma }\,,
 \label{kandPminimal}
\end{equation}
where $m_A$ are anti-Hermitian matrices:
$\delta _{\alpha \bar \beta }(m_A)^\alpha{}_\beta =-\delta _{\beta \bar\alpha  }(\overline{m}_A)^{\bar \alpha  }{}_{\bar \beta  }$.
Hence the gauge group is part
of U$(n)$.
This means that the compensator $X^0$ does not participate in Yang-Mills transformations, the zero component of the Killing vector vanishes, $k^0_A=0$,  and other components do not depend on $X^0$.

For the decoupling of matter from compensator a more general class of the
K{\"a}hler manifold for all $n+1$ chiral multiplets is possible. However,
more general choices after gauge fixing of the local conformal symmetry
will lead to noncanonical kinetic terms for scalars.  Our choice of
scalar-independent $\Re f_{AB} $ was made to have canonical kinetic terms
for vectors. Our choice of $\Im f_{AB}$ and of cubic superpotentials
could have included a dependence on some ratios of homogeneous scalars
${X^\alpha\over X^\beta} $. We do not consider such theories here for
simplicity, and also since they would blow up at $\langle X^\beta
\rangle=0$.

We impose  our 4 conditions above on  the superconformal action (\ref{sconfaction}) and find a superconformal action of this special kind.  The scalar-gravity part of the superconformal action (neglecting fermions and gauge vector fields)\footnote{The complete action for this class of models is presented  in the Appendix~\ref{app:completeAct}.}
becomes
 \begin{eqnarray}
 {1\over \sqrt{-g}}\hat { \mathcal{L}}_{\rm sc}=\ft16(|X^0|^2 - |X^\alpha|^2) R
-\eta_{I\bar J}D^\mu X^I\,D_\mu \bar X^{\bar J}-\delta^{\alpha\bar \beta}{\cal W}_\alpha \overline{{\cal W}}_{\bar \beta} - \ft12(\Re f)^{-1\,AB} {\cal P}_A{\cal P}_B\, .
 \label{scGravhat}
\end{eqnarray}
We may split the total superconformal action into parts depending on the compensator multiplet $X^0, \Omega^0, F^0$ and the part not depending on it. In our class of models  we get
 \begin{eqnarray}
{1\over \sqrt{-g}}\hat { \mathcal{L}}^{\rm 0}_{\rm sc}=\ft16 |X^0|^2 R + D^\mu X^0\,D_\mu \bar X^{\bar 0} \ ,
 \label{0}
\end{eqnarray}
 \begin{eqnarray}
{1\over \sqrt{-g}}\hat { \mathcal{L}}^{\rm m}_{\rm sc}=
-  \ft16 |X^\alpha|^2  R
-\delta_{\alpha \bar \beta}D^\mu X^\alpha\,D_\mu \bar X^{\bar \beta}-\delta^{\alpha \bar \beta}{\cal W}_\alpha  \overline{{\cal W}}_{\bar \beta} - \ft12(\Re f)^{-1\,AB} {\cal P}_A{\cal P}_B\, .
 \label{split}
\end{eqnarray}
Each of these two actions is separately superconformal (when fermions and vectors are added). In the absence of fermions and vectors they have local conformal and local $U(1)$ chiral symmetry. The matter part of the action $\hat { \mathcal{L}}^{\rm m}_{\rm sc}$ does not depend on $X^0$ and therefore   it remains superconformal after the gauge fixing.

\subsection{Gauge fixing}\label{gfix}

Now we proceed with the gauge fixing of  local symmetries that are absent in  supergravity.
 We change variables from the basis $\{X^I\}$ to
a basis  $\{y,\,z^\alpha\} $, where $\alpha =1,\ldots ,n$ using
$
X^I = y\, Z^I(z) $.
We now  fix
{\it the special conformal symmetry}:
\begin{equation}
 b_\mu =0\,.
\label{Kgauge}
\end{equation}
The {\it dilatational
 and $U(1)$ symmetries}  are fixed by a  choice $\mathcal{N}(X,\bar X)= -|X^0|^2 + |X^\alpha|^2= \Phi(z, \bar z)$ and\footnote{Here we restore the value of $M_P$ to stress that after the gauge fixing of the superconformal action only one dimensional parameter, $M_P$,  appears in the supergravity action. Moreover, in the class of models described above, in the Jordan frame, the matter part of the action does not depend on $M_P$, since it was independent on $X^0$.}
\be
X^0= \bar X^{\bar 0}=  \sqrt 3 \, M_P \,, \qquad  y=\bar y =1 \, , \qquad  X^\alpha = z^\alpha \ .
\label{Dgauge}\ee
The special supersymmetry is fixed\footnote{While the gauge fixing for dilatations agrees with the choice made in \cite{Ferrara:2010yw},
we made a different gauge choice for special supersymmetry, which is chosen here in order that the compensating
multiplet does not mix with the physical multiplets.} by the matching requirement on fermions
in which
\be
\Omega ^{0}=0\, ,  \qquad  \Omega ^{\alpha
}=\chi ^{\alpha } \,.
\ee
This choice of the gauge fixing provides a decoupling of the matter multiplets $( X^\alpha=z^{\alpha},  \Omega ^{\alpha
}=\chi ^{\alpha }, F^\alpha = \delta^{\alpha \bar \beta}  {\partial \overline W(\bar z)\over \partial \bar z^{\bar \beta}}) $ from the compensator multiplet $(X^0, \Omega^0, F^0)$.
This leads to  \begin{equation}
\hat  \Phi(z,{\bar z}) =-3 M_P^2 + \delta_{\alpha\bar\beta }z^{\alpha} \bar z^{\bar\beta } \ , \qquad {\cal W}(X) =W(z)=
  {1\over 3} d_{\alpha \beta \gamma} z^\alpha  z^\beta  z^\gamma \ .
\label{Turin-12}
\end{equation}
After the gauge fixing the scalar-gravity part of the supergravity action is
\begin{eqnarray}
{1\over \sqrt{-g}}\hat { \mathcal{L}}^{\rm 0}_{\rm sg}=\ft12 M_P^2 ( R +6 A_\mu \,A^\mu ) \ ,
 \label{0s}
\end{eqnarray}
 \begin{eqnarray}
{1\over \sqrt{-g}}\hat { \mathcal{L}}^{\rm m}_{\rm sg}=
-  \ft16 |z^\alpha|^2  R
-\delta_{\alpha \bar \beta}D^\mu z^\alpha\,D_\mu \bar z^{\bar \beta}-\delta^{\alpha \bar \beta} W_\alpha  \overline{{ W}}_{\bar \beta} - \ft12(\Re f)^{-1\,AB} P_AP_B\, ,
 \label{splits}
\end{eqnarray}
where the $U(1)$ $\mr$ covariant derivative acting on scalars is
\begin{eqnarray}
D_\mu z^\alpha = \partial_\mu z^\alpha -\rmi A_\mu
z^\alpha  \,.  \label{covder}
\end{eqnarray}
and
\be
W_\alpha\equiv  {\partial W(z)\over \partial z_\alpha}= d_{\alpha \beta \gamma}   z^\beta  z^\gamma \ .
 \ee
When the corresponding fermion and vector fields are added to this action,
one finds that the total action has an unbroken local Poincar{\'e} supersymmetry.
The crucial difference with a generic case of supergravity theory is that the matter part of action whose scalar-gravity part is given  in Eq. (\ref{splits}), remains superconformal invariant. In particular, the scalar-gravity action in Eq. (\ref{splits})  is invariant under simultaneous local conformal transformation of the metric and scalars, the  vector being inert,
\be
g_{\mu\nu} '= \rme^{-2\sigma(x)} g_{\mu\nu}\, , \qquad z'= \rme^{\sigma(x)} z \, , \qquad \bar z'= \rme^{\sigma(x)} \bar z
\, , \qquad A_\mu '= A_\mu \ .
\label{dil}\ee
It is also invariant under local $U(1)$ ${\mathcal{R}}$ symmetry, which is part of the superconformal $SU(2,2|1)$ symmetry,
\be
g_{\mu\nu} '=  g_{\mu\nu}\, , \qquad z'= \rme^{\rmi\Lambda(x) } z \, , \qquad \bar z'= \rme^{-\rmi\Lambda (x)} \bar z
\, , \qquad A_\mu '= A_\mu +\partial _\mu \Lambda(x) \ .
\label{U1}
\ee
 The action
(\ref{0s}) is nonconformal; it describes the gravitational multiplet, including
the auxiliary field $A_{\mu }$, and it is not invariant under local
conformal transformations nor under local $\mathcal{R}$-symmetry.
 The total self-coupling of the gravitational multiplet breaks the superconformal symmetry down to super-Poincar{\'e}.


 \subsection{A simple Jordan frame supergravity action with superconformal matter}\label{jframe}
 Let us summarize what we have learned in the previous couple of sections.
 A spectacular property of the supergravity total action in this class of models is its  local supersymmetry,
 whereas the kinetic terms are canonical and the potential is that of global SUSY. The total action
 can be  split into the action of the pure supergravity part and the superconformal matter part.  For example, the total scalar-gravity part of the supergravity action is
\begin{eqnarray}
\hat { \mathcal{L}}^{\rm 0}_{\rm sg}+\hat { \mathcal{L}}^{\rm m}_{\rm sc} = \sqrt{-g_J} \left[ \ft12 M_P^2 ( R +6 A_\mu \,A^\mu )-  \ft16 |z^\alpha|^2  R
-\delta_{\alpha \bar \beta} g^{\mu\nu} D_\mu z^\alpha\,D_\nu \bar z^{\bar \beta}-\hat V_J\right],
\label{L0}\end{eqnarray}
where
\be
\hat V_{J}=\delta^{\alpha \bar \beta} W_\alpha  \overline{{ W}}_{\bar \beta} + \ft12(\Re f)^{-1\,AB} {\cal P}_A{\cal P}_B \ .
\label{VJ0}\ee
It consists of a pure supergravity part to which one has to add the global supersymmetry action interacting with the supergravity multiplet.
If we would like to cut out the supergravity multiplet and get the global SUSY action, we would have to remove from the action the first 3 terms,
remove $\sqrt{-g}$, replace the curved metric by the flat one, and remove the $U(1)$ field $A_\mu$ from the covariant derivative. We would get
\begin{eqnarray}
\hat { \mathcal{L}}^{\rm m}_{\rm susy} =
-\delta_{\alpha \bar \beta} \eta^{\mu\nu} \partial_\mu z^\alpha\,\partial_\nu \bar z^{\bar \beta}-V_{\rm susy}  \ . \label{tot2}
\label{susy}\end{eqnarray}
The potential $V_{\rm susy} $ is precisely the same as in supergravity and given in Eq. (\ref{VJ0}).

Vice versa, if one would like to promote any scale-free global SUSY theory into  supergravity, one starts with the action (\ref{susy}) with the potential in Eq. (\ref{VJ0}). First, one has to add a factor $\sqrt{-g}$ to the global action and replace the flat space-time metric $\eta^{\mu\nu}$ in the kinetic term by the curved metric $g^{\mu\nu}$. The partial derivatives of the scalars have to be made ${\mathcal{R}}$ covariant
as in Eq. (\ref{covder})  to make the chiral multiplets superconformal:
\begin{eqnarray}
\hat { \mathcal{L}}^{\rm m}_{\rm susy} =
-\delta_{\alpha \bar \beta} \eta^{\mu\nu} \partial_\mu z^\alpha\,\partial_\nu \bar z^{\bar \beta}-V_{\rm susy}    \quad \Rightarrow \quad  \hat { \mathcal{L}}^{\rm m}_{\rm sc} = \sqrt{-g_J}( -  \ft16 |z^\alpha|^2  R
-\delta_{\alpha \bar \beta} g^{\mu\nu} D_\mu z^\alpha\,D_\nu \bar z^{\bar \beta}-\hat V_J)\,,
\label{promote}
\end{eqnarray}
where
\be
\hat V_{J}= V_{\rm susy}=\delta^{\alpha \bar \beta} W_\alpha  \overline{{ W}}_{\bar \beta} + \ft12(\Re f)^{-1\,AB} {\cal P}_A{\cal P}_B \, , \qquad D_\mu z^\alpha= \partial_\mu z^\alpha-\rmi A_\mu z^\alpha \ .
\label{hatV}\ee
It remains to add an action of pure supergravity $\sqrt{-g_J} \left[ \ft12 M_P^2 ( R +6 A_\mu \,A^\mu ) \right]$. The result is in Eq. (\ref{L0}).
 The same for vector multiplets. If we include fermions, the rules require also to introduce the interaction with  gravitino, as shown in Appendix A. The resulting action of the form
(\ref{L0}), (\ref{VJ0}) has a local super-Poincar{\'e} symmetry and the matter action has a superconformal symmetry. When fermions and vectors are included, the generalization of Eqs. (\ref{L0}) and (\ref{VJ0}) is given in
Appendix~\ref{app:completeAct}. The principle is the same, all chiral and vector multiplets start interacting with the gravitational Weyl supermultiplet, and the pure supergravity action is added.
 The total action, including fermions and vectors,  has local Poincar{\'e} supersymmetry.

If we would like to embed the scale-free global SUSY theory into supergravity in the Einstein frame, we would have to use the \K\, potential
\begin{equation}
\hat {\mathcal{K}}(z,{\bar z})=-3 M_P^2 \log \left(-\ft1{3 M_P^2}\hat \Phi(z,{\bar z})\right) = -3 M_P^2 \log \left( 1-\ft{1}{3 M_P^2}\delta _{\alpha\bar{
\alpha}}z^\alpha\bar z^{\bar{\alpha}}\,\right)  .
\end{equation}
The total scalar-gravity part of action will be
\begin{eqnarray}
\hat { \mathcal{L}}^{\rm E}_{\rm supergrav}= \sqrt{-g_E} \left( \ft12 M_P^2  R
-\hat {\mathcal{K}}_{\alpha \bar \beta} g^{\mu\nu}_E \partial _\mu z^\alpha\,\partial _\nu \bar z^{\bar \beta}- V_E\right), \label{tot4}
\end{eqnarray}
where $V_E$ is
\begin{equation}
V_{E}= \rme^{\mathcal{K}\over M_P^2}\left(
\nabla _\alpha Wg^{\alpha \bar\beta }\nabla _{\bar\beta }\overline{W}
-{3W\overline{W}\over M_P^2} \right ) +  \ft12(\Re
f)^{-1\,AB}P_AP_B\,, \label{VEbis}
\end{equation}
and  $P_A= -3 {{\cal P}_A \over \Phi(z, \bar z)}$.  The action in the Einstein frame is significantly different from the global SUSY action, the kinetic terms are not canonical, the $F$-term potential is complicated and no part of this action has a conformal or ${\mathcal{R}}$ symmetry. The dependence on $M_P$ is all over the place. Thus, for scale-free globally supersymmetric models there is an obvious advantage to study their supergravity embedding  in the simple Jordan frame with manifest superconformal symmetry of the matter action as shown in Eq. (\ref{L0}).
Note also that for this class of models the  potential in the Einstein frame can be given in the form
\be
 V_{E}= 9 {\hat V_{J}\over \Phi^2}=  {\delta^{\alpha \bar \beta} W_\alpha  \overline{{ W}}_{\bar \beta} + \ft12(\Re f)^{-1\,AB} {\cal P}_A{\cal P}_B \over ( 1 - \delta_{\gamma\bar\delta }z^{\gamma} \bar z^{\bar\delta }/3 M_P^2 )^2}\, ,
\ee
and it is positive semidefinite.

It is instructive to compare the CSS class of models with no-scale supergravity. The review of no-scale supergravity models can be found in
 Sec. XII in \cite{Ferrara:1987ju}. No-scale models  have a positive semidefinite potential in the Einstein frame. This condition is also satisfied by the CSS models. However, the second feature of no-scale models is that at the minimum  $V_{\rm min}=0$ they break supersymmetry spontaneously.
Meanwhile in the CSS models, the minimum of the potential is at $V=0$, but supersymmetry is not broken there. Therefore such theories may provide a natural starting point for investigation of the models with a low scale of SUSY breaking.

An interesting property of the no-scale supergravity is that the term $-3|W|^2$ is absent in the expression for the scalar potential. As we already demonstrated, the CSS models share this property, but, in addition, the expression for the scalar potential in the Jordan frame does not have the overall factor $\rme^{\cal K}$, the \K\, connection terms,  $K_\alpha W$,  drop and  $K^{\alpha \bar \beta}$ is replaced by $\delta^{\alpha \bar \beta}$. This is a major simplification, reducing the $F$-term potential to its global SUSY expression. We found this result directly from the superconformal approach to supergravity, but one can also confirm it by direct calculations presented in Appendix~\ref{app:Why}.

The important property of both CSS and no-scale supergravity models is
that in order to describe physics, we have to break some of the
symmetries of these models. In the case of the no-scale model an
important example is the KKLT stabilization of the string theory \K\,
moduli \cite{Kachru:2003aw} where the breaking of the no-scale property
of supergravity is achieved via the instanton corrections/gaugino
condensation. In the case of superconformal matter coupling we will
introduce in the next section a mechanism of breaking of superconformal
symmetry which is useful for inflation as well as for a possible solution
of the $\mu$-problem in the NMSSM. This mechanism is geometric: the
moduli space of chiral fields including the compensator field is not flat
anymore, but no dimensional parameters are introduced.


 \subsection{Breaking of superconformal  symmetry
 via $\chi$ terms: The real part of the holomorphic function in scalar-curvature coupling}\label{break}

An interesting possibility to break the superconformal symmetry of the matter multiplets in
the supergravity action without introducing dimensional parameters into the underlying superconformal action (\ref{sconfaction})
is to modify the real function $\mathcal{N}(X,\bar X)$ as follows:
\be
\mathcal{N}(X,\bar X)= -|X^0|^2 + |X^\alpha|^2 - \chi \, \left (a_{\alpha \beta} {X^\alpha X^{\beta} \bar X^{\bar 0}\over X^0}+ h.c.\right ) .
\label{J} \ee
Here $\chi$ is a dimensionless parameter and $a_{\alpha \beta}$ is a numerical matrix. The function  $\mathcal{N}(X,\bar X)$ has the correct dilatation weight  in each $X$ and $\bar X$ direction. This means that the new K{\"a}hler manifold for all $n+1$ chiral multiplets $X^I$, including the compensator field $X^0$, is not flat anymore. The metric
$
 G_{I\bar J}={\partial {\cal N}(X,\bar X)\over \partial X^I \partial \bar X^{\bar J} }
$
 is not flat and the curvature $R_{I\bar K J\bar L}$ is proportional to $ \chi$.
 We keep a cubic, $X^0$ independent superpotential and a flat vector moduli space and an $X^0, \bar X^{\bar 0}$ independent momentum map ${\cal P}_A$,  as above.
  The gauge fixing of this class of models with $\mathcal{N}(X,\bar X)=\Phi(z, \bar z)$ and $X^0=\bar X^{\bar 0}= \sqrt 3 M_P$ leads to a Jordan frame supergravity,
  described in the general case in \cite{Ferrara:2010yw}.  The resulting supergravity action in which the matter multiplet is not superconformal due to $\chi$ terms is given by
\begin{eqnarray}
 {1\over \sqrt{-g_J}} { \mathcal{L}}^{\rm J}_{\rm sg}=   \ft12 M_P^2 \left(R +6 A_\mu \,A^\mu \right)-  \ft16 \left ( |z^\alpha|^2 -  \chi( a_{\alpha \beta} z^\alpha z^\beta +h.c. )\right ) R
-\delta_{\alpha \bar \beta}  D_\mu z^\alpha\,D^\mu \bar z^{\bar \beta}-V_J \label{tot5} \ ,
\end{eqnarray}
where
\be
V_J=G^{\alpha \bar \beta}W_\alpha  \bar W_{\bar \beta} + \ft12(\Re f)^{-1\,AB} {\cal P}_A{\cal P}_B \ .
\label{VJ}
\ee
Here $G^{\alpha \bar \beta}$ is the matter part of the inverse  metric $G^{I\bar J}$ of the enlarged space including the
compensator. We compute it in Appendix ~\ref{app:G}.
The action corresponds to a Jordan frame supergravity with the frame function given in Eq. (\ref{Phi})
where the holomorphic function is $J(z)= -  \chi a_{\alpha \beta} z^\alpha z^\beta $. Note that the inverse metric $G^{\alpha \bar \beta}$ in the potential (\ref{VJ})
is not flat anymore; it depends on moduli. However, the kinetic term for scalars is canonical since  $\Phi _{\alpha\bar\beta }= \delta _{\alpha\bar\beta }$. An additional simplification is also observed in the potential  (\ref{VJ}) : it has the form rather close to the global supersymmetry potential. The difference comes from the nonflat inverse metric  $G^{\alpha \bar \beta}$.
In particular, {\it certain directions may still keep a flat metric and the corresponding part of the potential remains superconformal}. As we will see later, this property is useful for the
studies of inflation in the Jordan frame.

\subsection{A simple example}

A simple example is the case of two scalars, the field $S$, which is not
included in the $\chi$ term, and the field $H$, which is included in the
$\chi$ term. We start with the superconformal theory (\ref{J})
\begin{equation}
  \mathcal{N}(X,\bar X)= -|X^0|^2 + |S|^2 +|H|^2  -{3\over 4}  \chi \,
\left ( {H^2 \bar X^ 0 \over X^0}+ h.c.\right )\,.
 \label{J1}
\end{equation}
After
gauge fixing $X^0=\bar X^{\bar 0}= \sqrt 3 M_P$ we find a Jordan frame
supergravity with the frame function
\begin{eqnarray}
\Phi&=& -3M_P^2  + |S|^2 +|H|^2
-{3\over 4}  \chi \,  ( H^2 +\bar H^2 ) \nonumber\\ &=& -3 M_P^2  + |S|^2
- {1\over 4}\left (1+  {3\over  2}\chi \right)  (H-\bar H)^2  + {1\over
4}\left (1-  {3\over 2}\chi \right)  (H+\bar H)^2\,.
\label{AL}
\end{eqnarray}
The
action has the following curvature-dependent terms:
\begin{equation}
 \left [{1\over 2}
M_P^2  -{1\over 6}  |S|^2 -{1\over 6} |H|^2  + {1\over 8}  \chi \,  ( H^2
+\bar H^2 ) \right] R\,.
\label{Rcoupling}
\end{equation}
If $S=0$ and  the field $H$
is real, so that $H=\bar H={h\over \sqrt 2}$, we find the following
action:
\begin{equation}
  {1\over 2} \left [ M_P^2 +( -{1\over 6} + {1\over 4}  \chi
)\,  h^2 \right] R \,.
\end{equation}
This will explain a particular relation
between the standard model action \cite{Sha-1} and the NMSSM action
during inflation, as well as the relation $\xi= -{1\over 6} + {1\over 4}
\chi$; see  (\ref{ident}) in Sect. \ref{basicinfl}.

We may also rewrite the curvature-dependent terms of the action (\ref{Rcoupling}) in the following form:
\be
\left [{1\over 2} M_P^2  -{1\over 6} \left ( |S|^2 - {1\over 4}\left (1+   {3\over  2}\chi\right)  (H-\bar H)^2  + {1\over 4}\left (1-  {3\over 2}\chi\right)  (H+\bar H)^2 \right ) \right] R\ .
\label{newform}\ee

\subsection{Shift symmetric models}
\label{shift}

In the main part of the paper we will be interested in the regime where $\chi \gg 1$. However, there are two other special cases which may be equally interesting, $\chi=\pm {2\over 3}$:

1) $\chi=-{2\over 3}$, in which case the frame function is given by
\be
\Phi_{\chi=-2/3}= -3 M_P^2  + |S|^2   + {1\over 2}  (H+\bar H)^2  \ .
\ee
In this case the field $S$ remains conformally coupled, but the imaginary part of the field $H$, which is given by $(H-\bar H)/2i$,  decouples from the curvature scalar in (\ref{newform}), i.e. this field becomes minimally coupled.

2) $\chi={2\over 3}$,
\be
\Phi_{\chi=2/3}=-3 M_P^2  + |S|^2   - {1\over 2}  (H-\bar H)^2  \ .
\ee
In this case $S$  remains conformally coupled, but the real part of the field $H$, which is given by $(H+\bar H)/2$,  decouples from the curvature scalar in (\ref{newform}), i.e. it becomes minimally coupled.

Consider a particular class of superconformal models with the superconformal symmetry broken by the real part of the holomorphic function, as shown in (\ref{J}). For superconformal models we are interested in the relation between the frame function and the \K\, potential of the form
\begin{equation}
{\mathcal{K}}=-3 M_P^2 \log \left(-\ft1{3 M_P^2} \Phi(z,{\bar z})\right)  .
\end{equation}
If we break the superconformal symmetry of matter preserving this relation between the frame function and the \K\, potential, we are led to a class of models where the \K\, potential has a shift symmetry:

1) $\chi=-{2\over 3}$,
\begin{equation}
{\mathcal{K}}(z,{\bar z})_{\chi=-2/3}=-3 M_P^2 \log \left(1- {1\over 3M_P^2} \left  (|S|^2   + {1\over 2} (H+\bar H)^2\right )\right).
\end{equation}
This \K\, potential has a shift symmetry with respect to $H-\bar H$.

2) $\chi={2\over 3}$,
\begin{equation}
{\mathcal{K}}(z,{\bar z})_{\chi=2/3}=-3 M_P^2 \log \left(1- {1\over 3M_P^2} \left  (|S|^2   - {1\over 2} (H-\bar H)^2\right ) \right).
\end{equation}
This \K\, potential has a shift symmetry with respect to $H+\bar H$.

Thus, a new class of models with the shift symmetric  \K\, potential was
derived here from the superconformal approach to supergravity. These
models provide a natural basis for a broad class of new models of chaotic
inflation in supergravity, with a functional freedom of choice of the
inflaton potential  \cite{Kallosh:2010ug}.

\subsection{Stabilization of moduli at the origin of the moduli space}
\label{origin} We may be interested for applications in a method of
breaking superconformal symmetry which enforces  some scalars to be fixed
at the origin of the moduli space. The method to achieve the moduli
stabilization at the origin of the moduli space due to quartic
corrections to the \K\, potential, $\mathcal{K}(S,{\bar S})= S \bar S - {
(S\bar S)^2\over \Lambda^2}$, was studied in \cite{Kitano:2006wz}.  It
was argued there that the quartic term originates from the loop
corrections representing the effective potential  from the  massive
fields  which have been integrated out. The sign of the second term in
the \K\,  potential is negative. In such a case the supergravity potential
was shown in \cite{Kitano:2006wz} to stabilize at $S=0$, at the origin of
moduli space. Such quartic terms may be generated by radiative
corrections, or they may even be present in the \K\, potential from the
very beginning.

Here we will show how these terms may emerge  from superconformal coupling of matter if one introduces an additional  coupling of matter to the compensator.  We split the $n+1$ scalars into a group where $X^0$ is the compensator field,  $X^a=\{ X^1, ..., X^{n-1}\}$ are matter scalars and $X^n$  is the field which we would like to stabilize at the minimum of the potential at $X^n=0$.
We consider a superconformal theory where the $X^n$ direction is not present in the $a_{\alpha \beta}$-matrix:
\be
\mathcal{N}(X,\bar X)= -|X^0|^2 + |X^a|^2 +|X^n|^2- \chi \, \left (a_{ab} {X^a X^{b} \bar X^{\bar 0}\over X^0}+ h.c.\right )- 3 \zeta  {|X^n\overline {X^{ n}}|^2\over X^0\overline {X^{ 0}}}.
 \ee
After gauge fixing at $X^0=\overline {X^{ 0}}=\sqrt 3\, M_P$ and using  notation  $X^a= z^a$, $X^n=S$, we find the frame function
\be
\Phi(z^a, \bar z^b; S, \bar S)= -3 \rme^{-{1\over 3} \mk  (z^a, \bar z^b; S, \bar S)}=
 -3 M_P^2 + |z^a|^2 +|S|^2- \chi \, \left (a_{ab} z^a z^{b}+ h.c.\right )- \zeta  {|S\overline {S}|^2\over  M_P^2}.
 \ee
As we will find later, in agreement with the proposal in \cite{Lee:2010hj} and previous work in \cite{Kitano:2006wz}, the term $- \zeta  {|S\overline {S}|^2\over  M_P^2}$ will allow us to stabilize the inflationary trajectory in the NMSSM  at $S=0$. An interesting feature of this mechanism is that the term $- \zeta  {|S\overline {S}|^2\over  M_P^2}$ vanishes  on  the inflationary trajectory when the moduli stabilization is achieved.

\section{Supergravity embedding of the scale-free NMSSM}
 \label{ss:sugraNMSSM}
 \subsection{Superconformal embedding of the NMSSM into supergravity }
\label{SC-Embedding} The original motivation for the NMSSM model which,
in addition to  two charged Higgs doublets $H_u, H_d$,  has a  gauge
singlet Higgs field $S$,  was  the hope for the elegant solution of the
$\mu$-problem. In MSSM there is a problem to explain a small value of the
$\mu$ term in the quadratic part of the superpotential $W=\mu \, H_u\cdot
H_d$. This term is required for the phenomenological reasons. In presence
of the gauge singlet $S$ one can start with the cubic superpotential
$\lambda S H_u\cdot H_d$ and hope to find a way to produce a small vacuum expectation value (VEV) of
$S$ so that $\mu_{\rm eff}= \lambda \langle S \rangle $ will produce the
effective desired value of the $\mu$ term.

From the superconformal approach we have a totally different motivation for the gauge singlet field. The scale-free NMSSM has cubic potential. Without the gauge singlet the term $\lambda S H_u\cdot H_d$  would not be possible. So from our perspective the motivation for the gauge singlet $S$ is the requirement of a  scale invariance of a globally supersymmetric  theory, which permits a simple promotion to local supersymmetry with the superconformal matter-supergravity coupling.

We start with the scale-free NMSSM model reviewed most recently in \cite{Ellwanger:2009dp}. The Higgs field sector of the NMSSM gauge theory has one gauge singlet and two gauge
doublet chiral superfields
$
z_H=\left\{ S,H_{u},H_{d}\right\}
$.
\begin{eqnarray}
S\, , \qquad
H_u =\left(
\begin{array}{c}
H^+_{u} \\
H^0_{u}
\end{array}
\right)  \, , \qquad H_d=\left(
\begin{array}{c}
H^0_{d} \\
H_{d}^-
\end{array}
\right)\,.
\label{doublets}\end{eqnarray}
and
$
H_{u}\cdot H_{d}\equiv -H_{u}^{0}H_{d}^{0}+H_{u}^{+}H_{d}^{-}
$.
The Higgs part of the model depends on five
chiral superfields.
The
superpotential is
\begin{equation}
W_{\rm Higgs}=-\lambda SH_{u}\cdot H_{d}+{\rho\over 3} S^{3}.
\label{W}\end{equation}
The quarks and leptons $
z_{QL}=\left\{ Q, U_R, D_R, L, E_R \right\}
$ are introduced via Yukawa cubic superpotential $W_{\rm Yukawa}$ so that the total superpotential for all superfields $z^\alpha= (z_H, z_{QL})$ is cubic
\be
W_{\rm total} =W_{\rm Yukawa}+ W_{\rm Higgs} ={1\over 3} \, d_{\alpha \beta \gamma} \, z^\alpha z^\beta z^\gamma \ .
\ee
The $D$- and $F$-term potentials of the general form (\ref{VJ0}) for the
NMSSM are given explicitly in Eqs. (9) and (10) of  \cite{Franke:1995tc}
where also the complete set of Feynman rules is presented. All kinetic
terms are  canonical, both for chiral as well as vector superfields. Also
the Yukawa and vector parts of the action as well as interaction between
the chiral  and vector multiplets are given explicitly. We do not add the
soft breaking terms to the NMSSM  at this point since we would like first
to embed the globally supersymmetric action into supergravity.

We have shown above that the scale-invariant version of the NMSSM has all conditions satisfied so that the simplest possible embedding of the scale-invariant version of it  into supergravity is possible.  One should take the globally supersymmetric action of the form
(\ref{susy}) with details in  \cite{Franke:1995tc} and follow the rules explained around Eqs. (\ref{promote}) and (\ref{hatV}). This gives the promotion to supergravity of the scale-invariant globally supersymmetric NMSSM.

The full supergravity action corresponds to the choice of the frame function
\begin{equation}
   \hat \Phi (z,{\bar z})=  -3M_P^2 + (S \bar S + H_u H_u^\dagger + H_d H_d^\dagger) \, ,\label{SCframe}
\end{equation}
which corresponds to the underlying superconformal theory (\ref{scGrav}) with an extra compensator field $X^0$
\begin{equation}
   \mathcal{N}(X,\bar X)=  - X^0 \bar X^{\bar 0} + (S \bar S + H_u H_u^\dagger + H_d H_d^\dagger) \, .\label{SCframe1}
\end{equation}
The supergravity potential in the Jordan frame is the same as the global one given in  eqs. (9) and (10) of  \cite{Franke:1995tc}. For example, the Higgs-gravity part of the supergravity action consists of the supergravity part, given in Eq. (\ref{0s})
and the matter part of supergravity action, which is superconformal, when interacting with the Weyl multiplet:
\begin{eqnarray}
\hat { \mathcal{L}}^{\rm m}_{\rm sc} = \sqrt{-g_J} \left[ -  {R\over 6} \left(S \bar S + H_u H_u^\dagger + H_d H_d^\dagger\right)
- D_\mu H^u\,D^\mu H_u^\dagger- D_\mu H^d\,D^\mu H_d^\dagger - D_\mu S\,D^\mu S^\dagger  -\hat V_J\right]\, , \label{totNMSSM}
\label{NMSSM sugra1}
\end{eqnarray}
where
\be
\hat V_J= \left |{\partial W\over \partial S}\right |^2 + \left |{\partial W\over \partial H_u}\right |^2 + \left |{\partial W\over \partial H_d}\right |^2 +{\frac{g^{^{\prime }2}}{8}}(|H_{u}^{0}|^2-|H_{d}^{0}|^2)^2+{
\frac{g^2}{8}}(H_{u}^{\dagger }\vec{\tau}H_{u}+H_{d}^{\dagger }\vec{
\tau}H_{d})^2 \ .
\ee
Here $D_\mu$ acting on scalars includes  the gauge field of the $U(1)$ $\mr$ symmetry $A_\mu$,
which is an auxiliary field of supergravity; see Eq. (\ref{covder}).
It can be replaced by its on-shell value as the function of scalars and fermions; see e.g. Eq. (5.13) of \cite{Ferrara:2010yw}.
For example, its bosonic part $\mathcal{A}_\mu(z, \bar z)$ is given in Eq. (\ref{A}).
\subsection{Breaking superconformal symmetry of matter in the NMSSM supergravity }
\label{EJ-Embedding}

Here we consider a possibility to break the superconformal symmetry of the matter multiplets in supergravity action geometrically, without introducing dimensional parameters into the underlying superconformal action. One of the possibilities was studied in Sec. \ref{break}. It corresponds to  the choice of the frame function
\begin{equation}
   \Phi_\chi (z,{\bar z})=  -3M_P^2+ |S|^2 + |H_u|^2 + |H_d|^2 + \ft32 \chi (H_u\cdot  H_d + \hc)\, .\label{chi}
\end{equation}
This choice was proposed in \cite{Einhorn:2009bh} for the purpose of the Higgs-type inflation in the NMSSM.
The underlying superconformal action is defined by the function $\mathcal{N}_\chi (X,\bar X)$ which is homogeneous of first degree in both $X$ and $\bar X$:
\be
\mathcal{N}(X,\bar X)_\chi = -|X^0|^2 +  |S|^2 + |H_u|^2 + |H_d|^2 + \ft32 \chi \, \left (H_u\cdot  H_d  {\bar X^{\bar 0}\over X^0} + h.c.\right ) .
\label{Nchi} \ee
Note that there is no superconformal symmetry breaking in the $S$ direction of the moduli space, namely, the metric $G_{S\bar S} =1=G^{S\bar S}$ remains flat, decoupled from the compensator sector and from the $H_u$ and $H_d$ sectors. Meanwhile,  the moduli space of the Higgs doublets, $H_u$ and $H_d$, is mixed with the compensator field $X^0, \bar X^{\bar 0}$, and it is nonflat, with
$\chi $-dependent curvature.

The \K\, function of the enlarged space   (\ref{Nchi}) after the gauge fixing with $X^0=\bar X^{\bar 0} = \sqrt 3 M_P$  corresponds to the frame function (\ref{chi}). The bosonic part of the supergravity  action  is as before $
 \sqrt{-g_J} \left[ \ft12 M_P^2 ( R +6 A_\mu \,A^\mu )\right]
$
and the matter part of supergravity action, which is superconformal (up to terms with $\chi$), is
\begin{eqnarray}
 \sqrt{-g_J} \left[-  \ft16 \left(|S|^2 + |H_u|^2 + |H_d|^2) - \ft14 \chi (H_u\cdot  H_d + \hc)\right)  R
- |D_\mu H_u|^2- |D_\mu H_d|^2 - |D_\mu S|^2  -V_J \right]\,.
\label{NMSSM sugra2}\end{eqnarray}
In this case, as shown in Sec. \ref{break}, the $D$-term potential in the Jordan frame is the same as in the case $\chi =0$; however, the $F$-term potential, as given by (\ref{VJ})
 has a specific deviation from the quartic superconformal potential,  since the metric $G^{\alpha \bar \beta}$ is not flat at $\chi \neq
 0$:
\be
V_J= G^{\alpha \bar \beta}W_\alpha  \bar W_{\bar \beta} +{\frac{g^{{\prime}\,2}}{8}}(|H_{u}^{0}|^2-|H_{d}^{0}|^2)^2+{
\frac{g^2}{8}}(H_{u}^{\dagger }\vec{\tau}H_{u}+H_{d}^{\dagger }\vec{
\tau}H_{d})^2 \ .
\ee
The metric $G^{\alpha \bar \beta}$ is the part of the inverse $G^{I\bar J}$ to the $G_{I\bar J}={\partial {\cal N}(X,\bar X)\over \partial X^I \partial \bar X^{\bar J} }$ metric. It is easy to compute using Eq. (\ref{Nchi} ).
One may notice, using $W_{\rm Higgs}=-\lambda SH_{u}\cdot H_{d}+{\rho\over 3} S^{3}
$  that at $S=0$ the only contribution to the $F$-term potential comes from the term
\be
(V_J^F)|_{S=0}= {\partial W\over \partial S} G^{S \bar S } {\partial   \bar W\over \partial {\bar S}} = \lambda^2
 G^{S \bar S } |H_u\cdot H_d|^2\ .
\label{VJ1}\ee Since the field $S$ does not enter in the $\chi$ term, one
finds that $G^{S \bar S }=1$ and therefore even after this breaking of
superconformal symmetry the specific part of the potential remains
quartic. This plays an important role for inflation where the
inflationary trajectory is at $S=0$.

To embed the NMSSM gauge theory into the Einstein frame supergravity with the superconformal symmetry breaking explained above, we have to use the  \K\, potential
\begin{equation}
\mathcal{K}_\chi (z,{\bar z}) = -3\log \left[ 1-
\ft{1}{3}\left(S \bar S + H_u H_u^\dagger + H_d H_d^\dagger\right) -\ft12 \chi \bigl(H_u\cdot  H_d + \hc \bigr)\right ] .
\label{K}\end{equation}

 \section{Phenomenological aspects of the NMSSM}\label{phen}

Here we start with the current point of view on the NMSSM, and its
problems, following \cite{Ellwanger:2009dp}, where the globally
supersymmetric model is studied in presence of the terms breaking
supersymmetry softly, which originate from a hidden sector of the
theory. We afterwards discuss the issues of  the NMSSM  from the
superconformal symmetry approach that we find useful both for the
particle physics phenomenology as well as for cosmology.

One of the reasons to augment the MSSM by the gauge singlet Higgs field
$S$ and study the NMSSM  was that the superpotential of the MSSM
contained the term $\mu H_u\cdot H_d$. It is difficult to explain the
required smallness of this term. In the NMSSM, one may generate the $\mu$
term  as $-\lambda  \, \langle S\rangle \, H_u\cdot H_d$ from the
superpotential $W_{\lambda} =-\lambda  \, S\, H_u\cdot H_d$. The problem,
however,  is  to explain why one cannot add the term $\mu H_u\cdot H_d$
to the NMSSM. To address this problem, one may assume that the
superpotential of the NMSSM must be scale invariant. This requirement
forbids terms such as $\mu H_u\cdot H_d$, as well as the tadpole term
$\sim S$ and the term $\sim S^2$, and allows only the cubic
superpotential $W_{\rm Higgs}=\lambda S H_u\cdot H_d +{\rho\over 3} S^3$.

Scale invariance of the NMSSM superpotential allows its consistent
embedding into the CSS. From the top-down perspective,  this scale
invariance can be interpreted as a consequence of the original
superconformal symmetry, protected by the decoupling of the light fields
from the conformal compensator. However, scale  invariance of the
NMSSM superpotential may result in the cosmological domain problem,
which we are going to analyze now.

At low energies one usually considers adding to the global SUSY potential  the soft SUSY breaking terms. The soft terms are of two types. There are mass terms for each
Higgs,
\be
V_{\rm soft}^m= m^2_{H_u} |H_u|^2 +m^2_{H_d} |H_d|^2 +m^2_{S} |S|^2\, .  \label{soft}\ee
There are also terms related  to a superpotential contribution to the potential:
i.e. there is a coupling $A_\lambda$, $A_\rho$ for each cubic term in the superpotential, times the real part of the superpotential. In case of the NMSSM with the cubic superpotential they are
\be
V_{\rm {soft}}^{W} ( {\rm NMSSM}) =   A_\lambda \, \lambda \, S \, H_u\cdot H_d +   A_{\rho} \, {1\over 3} \rho\,  S^3 + h.c.
\label{Wsoft}\ee
A continuous  global ${\mathcal{R}}$ symmetry of the total potential, when each scalar transforms as $z'= z^{i\Lambda}, \bar z= \rme^{-\rmi\Lambda} \bar z$,  is broken down to a discrete one due to the $V_{\rm soft}^W(\rm NMSSM)$  term. Namely, (\ref{Wsoft}) is invariant under
$\Bbb{Z}_{3}$ symmetry:
\be
S'=  \rme^{2\pi \rmi  n\over 3}  S\, ,  \qquad H_u'=  \rme^{2\pi \rmi n\over 3}  H_u\, ,  \qquad H_d'=  \rme^{2\pi \rmi n\over 3}  H_d\, ,
\label{Z3}\ee
where $n\in \mathbb{Z}$ and we assume that $A_\lambda$ and $A_\rho$ are real. In such a case, the theory has domain walls created once the $\Bbb{Z}_{3}$ symmetry is spontaneously broken after a restoration of a symmetric phase in the hot early universe. This creates large anisotropies of the CMB and contradicts a successful nucleosynthesis.

An interesting role is played here by the local $U(1)$
${\mathcal{R}}$ symmetry, which is part of the superconformal $SU(2,2|1)$
symmetry  (and it is not included into the super-Poincar{\'e} symmetry). As
explained  in \cite{Ferrara:2010yw}, the $\chi$ term required for
inflation in the NMSSM must be a sum of holomorphic and antiholomorphic
terms to keep the Jordan frame kinetic terms canonical, $
 \Phi(z,{\bar z}) = -3\,  M_P^2 + \delta_{\alpha\bar\beta }z^{\alpha} \bar z^{\bar\beta }
+J(z) + \bar J(\bar z)$. These $J\left(
z\right) +\overline{J}\left( \overline{z}\right) $ terms
in the frame function and in the \K\, potential not only break the
continuous ${\mathcal{R}}$ symmetry, but also break the discrete
$\Bbb{Z}_{3}$ symmetry (\ref{Z3}). A study of the $\Bbb{Z}_{3}$ symmetry
breaking terms in the supergravity Einstein frame potential shows that
the symmetry breaking term is an order six operator $\sim \chi \,  {
\lambda^2 h^6 \over M_P^2}. $ According to \cite{Abel:1995wk}, this
amount of $\Bbb{Z}_{3}$ symmetry breaking may not be sufficient to make
the domain walls disappear before the nucleosynthesis. However, we have
to take into account that the $J(z) + \bar J(\bar z)$ terms in the  \K\,
potential  may change the soft breaking SUSY terms in the potential, in
presence of a hidden sector  \cite{Barbieri:1982eh,Giudice:1988yz,Shirai:2008qt}. This possibility was
proposed in \cite{Lee:2010hj}. Here we will present a more detailed investigation of this scenario.

Suppose that chiral superfields $z^\alpha= \{\phi^a, \varphi^i \}$ are split into an observable sector $\phi^a$ and the hidden sector $\varphi^i$. Whereas the observable fields have weak scale VEV's $\sim 10^{-16} M_P$, the hidden sector scalars  have a much larger scale, but they are still much smaller than $M_P$.
Therefore one may expect that at present $W = W_{\rm obs} + W_{\rm hid} \approx W_{\rm hid}$, and $\rme^{\mathcal{K}\over 2 M_P^2} \approx 1$. In what follows, we will assume that $W_{\rm obs}$ is cubic in $\phi^a$, but we will not specify the superpotential of the hidden sector.  Up to an irrelevant complex phase, the gravitino mass is given by
\begin{equation}
m_{3/2}=\rme^{\mathcal{K}\over 2 M_P^2} \frac{\langle W\rangle}{%
M_{P}^{2}}\approx \frac{\left\langle W_{\rm hid}\right\rangle
}{M_{P}^{2}}\, .
\label{gravitino}\end{equation}

In what follows, the discussion will proceed in the Jordan frame
supergravity since it makes the conceptual points very clear. We will
write the $\chi$ term, which we added to the \K\, potential, as the real
part of the holomorphic function $J(\phi)$, quadratic in fields from the
observable sector,  $J(\phi)=-\chi \, C_{a b} \phi^a \phi^{b} $. This
allows us to keep the Jordan frame kinetic terms for the observable
sector canonical and to have only a dimensionless superconformal symmetry
breaking parameter $\chi$: \be
 \mathcal{K}(z,{\bar z}) = -3 M_P^2\log \left[ 1-%
\frac{\phi^{a} \bar \phi_{a} }{3 M_P^2} -\frac{J(\phi) }{3 M_P^2} -\frac{\bar J(\bar \phi) }{3 M_P^2}
-\ldots \right]\,.
\label{chi1}\ee
Here $\ldots$ stands for the terms depending on the hidden sector superfields. For the values of the fields much smaller than the Planck scale we may  expand the logarithm in Eq.
(\ref{chi1}):
\be
 \mathcal{K}(z,{\bar z}) =
\phi^{a} \bar \phi_{a}  +J(\phi)  +\bar J(\bar \phi)  +\ldots\,.
\ee
Now we can use the \K\, invariance to switch to a different \K\, potential and superpotential,
\be
\mathcal{K}_{\rm eff} (z,{\bar z}) = \mathcal{K}(z,{\bar z})  -J(\phi) - \bar J(\bar \phi)+\ldots \, , \qquad   W_{\rm eff}  =W\rme^{J(\phi)/ M_P^2} \,.
\ee
The new \K\, potential is canonical, but the superpotential has a
correction,
\begin{equation}
\mathcal{K}_{\rm eff}\left(\phi,{\bar \phi}\right) =\phi^{a} \bar \phi_{a} \, , \qquad
 W_{\rm eff}\rightarrow W\rme^{J\left(\phi\right)  / M_{P}^{2}}\approx W+\frac{ \left\langle W_{\rm hid }
\right\rangle}{M_{P}^{2}}J\left(\phi\right) \approx W+m_{3/2}J\left(
\phi\right)\,.
\end{equation}
Here we took into account (\ref{gravitino}). In the specific case of the
NMSSM, where $J={3\over 2} \chi H_u \cdot H_d$, one finds
\cite{Lee:2010hj}
\begin{equation}
  W_{\rm eff}= - \lambda \, S \, H_u\cdot  H_d +
{\rho\over 3} S^3 + {3\over 2} \, \chi \, m_{3/2} H_u\cdot  H_d \,.
 \label{softmu}
\end{equation}
Thus, the mere existence of the real part of the
holomorphic  quadratic correction to the frame function for observable
Higgs fields, breaking the superconformal symmetry in a way required for
inflation, is responsible also for the specific contribution ${3\over 2}
\, \chi \, m_{3/2} H_u\cdot  H_d$ to the $\mu$ term in the effective
superpotential for small fields,
\begin{equation}
  \mu_{\rm eff}  =\frac{3}{2}\chi
m_{3/2}-\lambda \left\langle S\right\rangle \,.
\end{equation}
This is a specific
realization of the Giudice-Masiero mechanism \cite{Giudice:1988yz}. Note
that the term ${3\over 2} \, \chi \, m_{3/2} H_u\cdot  H_d$ breaks the
$\Bbb{Z}_{3}$ symmetry of the real part of the scale-invariant
superpotential. To evaluate the significance of this effect, one may
estimate the correction to the soft breaking part of the potential
originating from the term ${3\over 2} \, \chi \, m_{3/2} H_u\cdot  H_d$:
\begin{eqnarray}
V_{\rm soft}^{W_{\rm eff}} =A_{\lambda }\, \lambda SH_{u}\cdot H_{d}+A_{\rho }\, \frac{%
\rho }{3}S^{3}+B_\mu \, \mu_{\rm eff} \, H_{u}\cdot H_{d}+h.c. \,.
\end{eqnarray}
This term contains the $\Bbb{Z}_{3}$-noninvariant term
\be
\Delta V =   \frac{3}{2} B_\mu\, \chi m_{3/2} \,  (H_u\cdot  H_d + h.c.)\,.
\ee
According to \cite{Abel:1995wk},  $\Bbb{Z}_{3}$ symmetry does not lead to the cosmological domain wall problem if the difference in vacuum energy between the different vacua separated by the domain walls is greater than $10^{-7}{v\over M_p} v^4 \sim 10^{-25} v^4$. We may now compare the potential energy for two vacua, which are degenerate for $\chi =0$.
Consider  ${z\over |z|} = \rme^{{2\pi \rmi   \over 3}n}$ and take one vacuum with $n=0$ and another one with $n=1$. For $B_\mu \sim \chi m_{3/2} \sim v$,  the energy difference is  $\sim \frac{3}{2} B_\mu \,  \chi m_{3/2}\,  v^2\sim v^4 $, which is  many orders of magnitude greater than the energy separation  required for the absence of domain walls.

One may wonder, whether all of these nice properties will be spoiled by
the tadpole problem? Indeed,  in generic models interactions with heavy
particles may induce large terms linear in $S$ in the superpotential, see
e.g. \cite{Nilles:1982mp,Lahanas:1982bk,Ferrara:1982ke,Abel:1995wk}.
Fortunately, this problem can be solved under certain conditions, as
explained in \cite{Ellwanger:2009dp}. In particular, in the theories with
$\mathcal{R}$ symmetry \cite{Abel:1996cr} a solution to the tadpole problem was
suggested. We believe that this solution applies to our model. Some other
proposals of how to stabilize the singlets in supergravity and avoid domain
walls can be found in \cite{Kolda:1998rm}.

Let us summarize our approach to the NMSSM phenomenology.

1) There are several different versions of the NMSSM, and many
inequivalent ways to incorporate each of these versions into
supergravity. We propose to incorporate the NMSSM into a CSS. This singles out the scale-invariant
version of the NMSSM. In general, the embedding of a global SUSY model can be
quite complicated, but the embedding of the NMSSM into the CSS is a
trivial exercise in the Jordan frame: one simply replaces usual
derivatives by covariant derivatives. The resulting theory has
superconformal symmetry, and all kinetic terms are canonical. This is a
unique property of the CSS approach, not shared by other methods of
embedding of the NMSSM into supergravity.

2) After the embedding, all fields in the NMSSM are massless. Then one
introduces masses due to gravitational effects and interaction with
hidden sector. This explains the smallness of all masses in the NMSSM as
a consequence of the underlying superconformal symmetry.

3) Adding the $\chi$ term  to the \K\, potential is equivalent to adding
a nonminimal coupling of the Higgs field to gravity, which is consistent
with our ideology of breaking the superconformal symmetry by
gravitational effects. Whereas the $\chi$ term was added in order to
realize Higgs inflation, it plays an additional role: it leads to a
specific  realization of the Giudice-Masiero mechanism of generation of
the $\mu$ term in the NMSSM \cite{Giudice:1988yz}. This mechanism breaks
$\Bbb{Z}_{3}$ symmetry and resolves the domain wall problem in the NMSSM,
whereas the tadpole problem may be solved due to ${\cal R}$ symmetry of
our construction.

\section{On Higgs-type inflation with nonminimal coupling in standard model}
\label{ss:Higgstypenonmin}
\subsection{Basic model}
Here we review the Higgs-type inflation with nonminimal scalar-curvature
$\protect\xi$ coupling studied in \cite{Sha-1}. We will focus on three
different ranges of the Higgs field VEV's, at the beginning of the last
60 e-foldings, at the exit from inflation, and at the present values of
the SM  Higgs. In \cite{Sha-1} the SM potential with canonical kinetic
term for the Higgs field $h$ is coupled to a gravitational field in the
Jordan frame:
\begin{equation}
\mathcal{L}_{J}=\sqrt{-g_{J}}\left[ {\frac{M^2+\xi h^2}2
}R\left( g_{J}\right) -\ft12\partial _\mu h\partial _\nu hg_{J}^{\mu\nu}-{\frac{\lambda
}{4}}(h^2-v^2)^2\right] \,. \label{Trn-1}
\end{equation}
At present, $h = v \sim 10^{-16} M_P$, and $M_P^2 = M^2 + \xi v^2$. Since $v$ is extremely small, we will ignore it in our investigation, and take $M= M_P =1$. The frame function for the action (\ref{Trn-1}) in this approximation  is
$
\Phi =-3 (1+\xi h^2)$ and the rescaling of the metric function $\Omega^2={1+\xi h^2}$ and the action can be rewritten as
\begin{equation}
\mathcal{L}_{J}=\sqrt{-g_{J}}\left[ {\frac{1+\xi h^2}2
}R\left( g_{J}\right) -\ft12\partial _\mu h\partial _\nu hg_{J}^{\mu\nu}-{\frac{\lambda
}{4}} h^4 \right] \,. \label{shap}
\end{equation}
In the  Einstein frame the action is
\begin{equation}
\mathcal{L}_E=\sqrt{-g_{E}}\left( \ft12 R(g_{E})-\ft12\partial _\mu \psi \partial _\nu\psi\, g_{E}^{\mu\nu}-U(\psi )\right) ,
\end{equation}
where
\begin{equation}
U(\psi )={\frac{\lambda }{4}}\left( {\frac{h^2(\psi
)-v^2}{{ {1 + \xi h(\psi )^2}}}}\right)^2 \label{U} ,
\end{equation}
and $\psi $ is a canonically normalized scalar in the Einstein frame, defined by
\begin{equation}
\rmd\psi \equiv  \rmd h \sqrt {\Omega^2 + 6 \xi^2 h^2\over \Omega^4}\,.
\end{equation}
A solution of this equation is
\begin{equation}
\psi = \sqrt{1 + 6\xi^{-1}}\, {\rm ArcSinh} \bigl(\sqrt{\xi+6\xi^2}\, h\bigr)  -\sqrt 6\, {\rm ArcTanh} \left({\sqrt 6\, \xi h\over \sqrt{1+ \xi h^2 + 6 \xi^2 h^2}}\right) \,.
\end{equation}
It is useful to present this solution in a simpler, asymptotic form for three different ranges of $h$.

(1) In the interval $0< h\ll \frac{1}{\xi }$ one has
\begin{equation}
\psi \approx h \ , \qquad U(\psi )\approx {\frac{\lambda }{4}}\psi^4 \,.\label{region1}
\end{equation}

(2) In the interval ${\frac{1}{{\xi }}} \ll h\ll {\frac{1}{\sqrt{\xi }}}$ one has
\begin{equation}
\psi \approx \sqrt{3\over 2}\, \xi h^2 \ , \qquad U(\psi )  \approx \
{\frac{\lambda }{6\xi^2}}\left({\psi\over 1+\sqrt{2\over 3} \psi}\right)^2 \,. \label{region2}
\end{equation}
At the upper part of this interval one has $\psi = O(1)$. The existence of this intermediate range was not taken into account in many recent papers on Higgs inflation. It will play an important role in our discussion of the unitarity bound in the next section.

(3) Finally, for $h \gg {\frac{1}{\sqrt{\xi }}}$ (or, equivalently, $\psi \gg 1$) one has
\begin{equation}
\psi \approx \sqrt{3\over 2}\, \ln\, (\xi h^2)  \ , \qquad U(\psi )  \approx
{
\frac{\lambda }{4\xi^2}}\left( 1+\rme^{-{\frac{2\psi }{
\sqrt{6}}}}\right)^{-2}  \label{Trn-2} \,.
\end{equation}
In this regime,  the potential in the Einstein frame is very flat, which leads to inflation.
As one can see from  (\ref{Trn-2}), the constant ($\psi $-independent) term in
the potential $U\left(\psi \right)$ is $\frac{\lambda}{4\xi^2}$, so nothing would work without the
nonminimal scalar-curvature coupling proportional to $\xi$.

The slow-roll parameters, for $\xi h^2 \gg 1$, are
\begin{eqnarray}
  \label{eq:7}
  \epsilon &
  \simeq&\frac{4  }{3
   \xi^2h^4}
  \;, \\
  \eta  &
  \simeq& -\frac{4}{3 \xi h^2 } \,.
\end{eqnarray}
Slow roll ends when $\epsilon\simeq 1$, so the field value at the end of
inflation is
$h_{\mathrm{end}}\simeq(4/3)^{1/4}/\sqrt{\xi}\simeq 1/\sqrt{\xi}$.
The number of e-foldings $N \gg 1$ during the slow roll of the field $h$ from its initial value $h_0$ is given by
\begin{equation}
  \label{eq:8}
  N   \simeq \frac{3}{4}\xi h_0^2
  \,.
\end{equation}
For a particular case $N \sim 60$, the amplitude of scalar perturbations of metric corresponds to the COBE (Cosmic Background Explorer) normalization for
\begin{equation}
  \label{eq:9}
  {\xi \over \sqrt{\lambda}}
  \simeq  5\times 10^4
  \,.
\end{equation}
The Hubble constant during inflation in this model is $H \approx {\sqrt{\lambda\over 3}}\, {1\over 2\xi}$.

\subsection{The unitarity bound?}\label{unitarity}

Recently several authors argued that one cannot rely on the description
of various processes in the  Higgs inflation model on an energy scale
exceeding the unitarity bound $\Lambda \sim 1/\xi$
\cite{Burgess:2009ea,Barbon:2009ya,Hertzberg:2010dc,Germani:2010gm}. For
the nonsupersymmetric standard model described above, with $\lambda =
O(1)$, this bound is dangerously close to the Hubble constant during
inflation $H \approx {\sqrt{\lambda\over 3}}\, {1\over 2\xi}$. In the
NMSSM one may consider the regime with $\lambda \ll 1$, where the
concerns about the unitarity bound do not seem to appear
\cite{Lee:2010hj}. This can be done by using the rescaling of several
parameters of the model.

Indeed, one can easily check that all observational consequences of the inflationary model described above, including the value of the potential, the Hubble constant,  the slow-roll parameters, the number of e-folds of inflation, the amplitude of scalar perturbations of metric, the spectral index $n_s$, and the ratio of tensor perturbations to scalar perturbations  $r$,
depend on only two combinations of parameters: $\xi h^2$ and ${\lambda\over \xi^2}$. Therefore all observational consequences of this model are invariant with respect to the simultaneous rescaling $\lambda \to c^2\lambda$, $\xi \to c\,\xi$ and $h\to h/\sqrt c$.  This means that one can study the inflationary regime for $\lambda =1$, $\xi  \simeq  5\times 10^4$, and then rescale it to smaller values of $\lambda$ to avoid the problems with the unitarity bound.

It is good to know that we have this possibility. However, whereas it is possible to use small $\lambda$ in the NMSSM, one cannot do it in the original nonsupersymmetric model of \cite{Sha-1}. Therefore it would be interesting to double-check whether one should worry about the unitarity bound in general.

Most of the arguments suggesting the existence of this bound are based on the investigation of the theory in the small field approximation $\psi \approx h$, where one can use an expansion $\psi  = h (1 + \xi^2 h^2+...)$.
For example, Ref. \cite{Barbon:2009ya} considers the potential (\ref{U}) at small values of the field $\psi$ where the potential can be expanded in powers of $\psi$ as
\begin{equation}
U(\psi )  _{\psi \rightarrow 0 } \;  \Rightarrow \;
{\lambda \over 4}  \psi^4(1 - 4 {\xi^2 \psi^2}+ O(({\xi^2 \psi^2})^2)+...
\end{equation}\label{smallh}
One may consider the term $- \lambda  {\xi^2 \psi^6}$, take two of the fields $\psi$, form a loop and integrate. This will produce a term proportional to $ \lambda  {\xi^2 \Lambda ^2 \psi^4}$, where $\Lambda$ is a cutoff. Repeating this step for all higher order terms, one may come to a conclusion that quantum corrections to ${\lambda \over 4}  \psi^4$ become uncontrollable if $\Lambda  > 1/\xi$.

However, it was suggested in \cite{Lerner:2009na} that ``the apparent generation of the new physics is an artifact of considering only two terms of the expansion when all terms are important''. For example, one-loop quantum corrections to the scalar potential involve knowledge of the scalar propagator in an external classical field $\psi$, which is equivalent to a
resummation of diagrams with an arbitrary number of external lines of the scalar field. One-loop corrections to the potential are proportional to $(U''(\psi))^2 \ln |U''(\psi)|$. Therefore these corrections during inflation are suppressed by an extra power of $\lambda\over \xi^2$, as well as by the asymptotic flatness of the potential (\ref{Trn-2}). Here we would like to take another look at this issue, and give an independent argument, which can be applied not only to the scalar potential, but also for kinetic terms and scattering amplitudes.

The key observation used in the derivation of the unitarity bound was that for $h \ll 1/\xi$, the expansion of the potential contains powers of $({\xi^2 \psi^2})^{2n}$. Replacing the operators $\psi^2$ by $\Lambda^2$ results in quantum corrections containing powers of ${\xi^2 \Lambda^2}$, and, consequently, to the estimate for the energy cutoff $\Lambda \sim 1/\xi$. However, this is true only if one is interested in quantum effects at very small values of the Higgs field,  $h \ll 1/\xi$, which is very far from the inflationary region $h \gtrsim 1/\sqrt\xi$.

As we already mentioned, for $h \gg 1/\xi$ the expansion of the potential in powers of $\psi$ is dramatically different. Indeed,  expansion of the potential $U(\psi)$ (\ref{region2})  in powers of $\psi$ in the intermediate range ${\frac{1}{{\xi }}} \ll h\ll {\frac{1}{\sqrt{\xi }}}$ does not contain the dangerous factors $({\xi^2 \psi^2})^{2n}$\, :
\begin{equation}
U(\psi )  \approx \
{\frac{\lambda }{6\xi^2}}\left({\psi\over 1+\sqrt{2\over 3} \psi}\right)^2 = {\frac{\lambda }{6\xi^2}} \left[\psi^2 -2\sqrt{2\over 3}\psi^3 +...\right].
\label{mediumh1}\end{equation}
 The dependence on $\xi$ in this expression is extracted into a single overall coefficient ${\frac{\lambda }{6\xi^2}}$, and all terms in the expansion are proportional to $\psi^n$. This is very much different from the small field regime, where the higher order terms were proportional to $({\xi^2 \psi^2})^{2n}$. To estimate how vulnerable Eq. (\ref{mediumh1}) could be with respect to quantum corrections, one may again replace some of the operators $\psi^2$ in this expansion by $\Lambda^2$. One can easily see that the higher order corrections will remain small for $\Lambda \ll \psi$.  At the lower boundary of the range ${\frac{1}{{\xi }}} \ll h\ll {\frac{1}{\sqrt{\xi }}}$, this leads to the same bound as before: $\Lambda \sim 1/\xi$.  However, at the upper boundary one has $\psi = O(1)$, which means that quantum corrections are not expected to be important until one reaches super-Planckian energies, which are well above the energy scale of inflation.

One can reach similar conclusions for quantum corrections during inflation, when $\psi > 1$ and the potential is given by  Eq. (\ref{Trn-2}). This means that the typical energy scale of inflation, $H \sim \sqrt\lambda/\xi$, is many orders of magnitude below the UV cutoff during this process. Of course, for the processes which occur long after inflation, when $h \approx \psi < 1/\xi$, the unitarity bound  will be much smaller, $\Lambda \sim 1/\xi$  \cite{Burgess:2009ea,Barbon:2009ya,Hertzberg:2010dc}, but this does not affect our ability to describe physical processes during inflation.

An attempt to derive the unitarity bound without using the small field
approximation was made in \cite{Germani:2010gm}. The authors considered
interaction of the inflaton field with gravity in the Jordan frame and
argued that the scattering amplitude $2h\to 2h$ exceeds the unitarity
bound at energy $E > 1/\xi$. However, the estimates made in
\cite{Germani:2010gm} ignored the nondiagonal kinetic terms mixing the
scalar field with gravity in the Jordan frame. These terms disappear in
the Einstein frame, and the estimate of the corresponding $2h\to 2h$
scattering amplitude shows that it does not violate the unitarity bound
at sub-Planckian energies.

In   \cite{Hertzberg:2010dc,Burgess:2009ea} it was argued that
investigation of scattering  of scalar particles on other scalar and
vector particles also gives rise to the unitarity bound $\Lambda \sim
1/\xi$. Once again, the calculations in
\cite{Hertzberg:2010dc,Burgess:2009ea} are based on the expansion $\psi =
h (1 + \xi^2 h^2+...)$, which is valid only for $h < 1/\xi$. In the most
interesting interval of the values of the Higgs field $h \gg 1/\xi$, one
can repeat the arguments given above and again come to the conclusion
that the higher order corrections are suppressed for $\Lambda < \psi$.
The authors of Ref. \cite{Bezrukov:2010jz} mentioned a possibility of the
unitarity cutoff $\Lambda \sim 1/\sqrt \xi$, but for $\xi \sim 10^{4}$
this cutoff is 2 orders of magnitude higher than the Hubble constant
during inflation, so it is harmless.

Recently, a consistent UV completion of the Higgs inflation model was
proposed in \cite{Giudice:2010ka}. Their scenario does not have any
problems at high energy, but it is more complicated. For a proper choice
of the coupling constants, its inflationary predictions coincide with the
predictions of the original model of Ref. \cite{Sha-1}.

In conclusion, we do not think that one should worry too much about the
unitarity bound during inflation in the Higgs inflation model of Ref.
\cite{Sha-1}. However, those who want to feel even better protected
against this problem may either try to use a consistent UV completion of
this model \cite{Giudice:2010ka}, or switch to the NMSSM and study the
model with $\lambda \ll 1$, where the presumed unitarity bound $\Lambda
\sim 1/\xi$ is well above the typical  energy scale of inflation $H
\approx {\sqrt{\lambda\over 3}}\, {1\over 2\xi}$  \cite{Lee:2010hj}. That
is what we are going to do now.

\section{Inflation in the NMSSM}
 \label{ss:InflsugraNMSSM}

Here we will start with the Jordan frame supergravity (we set $M_{P}=1$
throughout this section) with the following frame function, as outlined in Secs. \ref{break}, \ref{origin} :
\begin{equation}
   \Phi (z,{\bar z})=  -3+ (S \bar S + H_u H_u^\dagger + H_d H_d^\dagger) +\ft32 \chi (H_u\cdot  H_d + \hc)- \zeta (S\bar S)^2\, .\label{Ste-1}
\end{equation}
Here the term $(S \bar S + H_u H_u^\dagger + H_d H_d^\dagger)$
corresponds to the superconformal  coupling of the chiral multiplets. The
term $+\ft32 \chi (H_u\cdot  H_d + \hc)$ is the  real part of the
holomorphic quadratic function in the curvature-scalar coupling; it
breaks the superconformal symmetry of the chiral multiplet coupling. This
term reflects the need of the superconformal symmetry breaking to provide
a realistic Higgs-type inflationary model proposed in
\cite{Einhorn:2009bh} and  developed in \cite{Ferrara:2010yw,Lee:2010hj}.
Finally, the term $\zeta (S\bar S)^2$ is added to provide the stability
of the origin of the moduli space, $S=0$, as proposed in
\cite{Lee:2010hj} and used in earlier models in \cite{Kitano:2006wz}.

To embed the NMSSM gauge theory into the Einstein frame supergravity we
will use the  \K\, potential and the superpotential
\begin{equation}
\mathcal{K}(z,{\bar z}) =-3 \log (-{1\over 3} \Phi) = -3\log \left[ 1-
\frac{1}{3}\left((S \bar S + H_u H_u^\dagger + H_d H_d^\dagger)\right) -{ \chi\over 2} (H_u\cdot  H_d + \hc)+ {\zeta\over 3} (S\bar S)^2\right] .
\label{K1}\end{equation}
\begin{equation}
  W =-\lambda SH_{u}\cdot H_{d}+{\rho\over 3} S^{3}\,,
\end{equation}
where the Higgs doublets are defined in (\ref{doublets}).
Note that
\begin{equation}
H_{u}\cdot H_{d}\equiv -H_{u}^{0}H_{d}^{0}+H_{u}^{+}H_{d}^{-}.
\end{equation}
\begin{equation}
  H_u H_u^\dagger + H_d H_d^\dagger= H_u^0 (H_u^0)^\dagger + H_d^0 (H_d^0)^\dagger +H_u^+ (H_u^+)^\dagger + H_d^- (H_d^-)^\dagger\,.
\end{equation}
As in \cite{Ferrara:2010yw}, we use a consistent truncation in which the charged
superfields $ H_{u}^{+}$ and $H_{d}^{-}$ are absent.  We will present later in Appendix
\ref{chargemass} the condition for the stability of the inflationary
trajectory  with regard to the vanishing charged fields. We will  use the
fact that the dependence on the neutral and charged Higgs fields in $H_u
H_u^\dagger + H_d H_d^\dagger$ is symmetric, whereas the one in
$H_{u}\cdot H_{d}$ is antisymmetric.

Below we use a simplified action of the NMSSM, containing only three
superfields: $S$, $ H_{u}^{0}$ and $H_{d}^{0}$, such that,
\begin{equation}
H_{1}=\left(
\begin{array}{c}
0 \\
H_{u}^{0}
\end{array}
\right) \,,\qquad H_2=\left(
\begin{array}{c}
H_{d}^{0} \\
0
\end{array}
\right) \,.
\end{equation}
With this truncation, the frame function, the   \K\, potential and the superpotential
are:
\begin{eqnarray}
{\textstyle}\Phi (z,{\bar z}) &=&-3+{\textstyle}\left(
|S|^2+|H_{u}^{0}|^2+|H_{d}^{0}|^2\right)
-\ft32\chi
(H_{u}^{0}H_{d}^{0}+\overline{H_{u}^{0}}\overline{H_{d}^{0}}) -\zeta \left|
S\right| ^{4}\,,
\label{Phi1}\end{eqnarray}
\begin{equation}
\mathcal{K}(z,{\bar z}) = -3\log \left[ 1-%
\ft{1}{3}\left( \left| S\right| ^{2}+\left| H_{u}^{0}\right| ^{2}+\left|
H_{d}^{0}\right| ^{2}\right) +\ft{1}{2}\chi \left( H_{u}^{0}H_{d}^{0}+%
\overline{H_{u}^{0}}\overline{H_{d}^{0}}\right) +\ft{1}{3}\zeta \left|
S\right| ^{4}\right] ,
 \label{K-NMSSM}\end{equation}
\begin{eqnarray}
W &=&\lambda SH_{u}^{0}H_{d}^{0}+{\rho\over 3} S^{3}\,.  \label{W-NMSSM}
\end{eqnarray}
The $D$-term potential in the Jordan frame remains simple
\begin{equation}
V_{J}^{D}={\frac{g^{^{\prime }2}}{8}}(|H_{u}^{0}|^2-|H_{d}^{0}|^2)^2+{
\frac{g^2}{8}}(H_{u}^{\dagger }\vec{\tau}H_{u}+H_{d}^{\dagger }\vec{
\tau}H_{d})^2\, .  \label{Ste-5}
\end{equation}
The $S$-dependent terms in the $F$-term potential, even in the Jordan frame, are complicated due to $\zeta$ corrections. However, we will establish that the stabilization of some scalars takes place and only one real scalar remains light during inflation. We will find out that during inflation all complicated corrections to the potential drop and we can explain the inflationary dynamics regime using the simple features of the superconformal matter coupling and its particular breaking.

 \subsection{Basic features of inflation in the NMSSM}\label{basicinfl}

For a numerical investigation of inflation in the NMSSM model with three
chiral multiplets and truncated charged Higgs fields we use the
Mathematica code \cite{Kallosh:2004rs} designed to compute the Einstein
frame potentials and scalar kinetic terms for any number of moduli with
generic \K \, potential ${\cal K}(z, \bar z)$ and generic superpotential
$W(z)$.

The potential in the NMSSM depends on three complex superfields:
\begin{equation}
S=s\rme^{\rmi\alpha }/\sqrt 2\,,\qquad H_{u}^{0}=h_{1}\rme^{
\rmi\alpha _{1}}/\sqrt 2\,,\qquad
H_{d}^{0}=h_2\rme^{\rmi\alpha _2}/\sqrt 2\,.
\label{complex}
\end{equation}
Note that here we slightly deviate from the notation of our previous
paper \cite{Ferrara:2010yw}: We divided all fields by $\sqrt 2$. The main
reason to do it is  to keep the fields $h$ canonically normalized in the
Jordan frame. It will simplify the comparison of inflation in the NMSSM
with inflation in the nonsupersymmetric standard model  \cite{Sha-1}.

The standard mixing of the Higgs fields is defined as
\begin{equation}
h_{1}\equiv h\cos \beta \,,\qquad h_2\equiv h\sin \beta \,,
\label{ansatz}
\end{equation}
which leaves us with two real fields, $h$ and $\beta $, instead of
$h_{1}$ and $h_2$. The $D$-flat direction, defined by
$
V_{J}^{D}=0
$
requires that
\begin{equation}
\beta =\pi/4;~~~~h_{1}^2=h_2^2=h^2/2.  \label{Ste-9}
\end{equation}

In this section we will consider the simplest inflationary solution with
$\beta = \pi/4$, $\alpha_i = 0$ and $s = 0$. In the next sections we will
investigate the conditions required for stability of this solution with
respect to the $\beta$, $\alpha_i$, and $s$.

We find that in the Jordan frame the total supergravity action for the
field $h$, under the condition that $\beta = \pi/4$, $\alpha_i = 0$, and
$s = 0$,  is reduced to
\begin{equation}
\mathcal{L}_{J}(h, g_J; \lambda)=\sqrt{-g_{J}}\left[ {1\over 2}\left( 1-{
\textstyle\frac{1}{6}} h^2
+\ft14\chi h^2 \right) R(g_{J}) -\ft12(\partial_\mu h)^2- {\lambda^2\over 16} h^4
\right] \, .
\label{inflation}\end{equation}

An interesting question to ask here is: {\it why} the complete
supergravity action of the NMSSM model with the frame function in
(\ref{Phi1}) and superpotential in (\ref{W-NMSSM}) at the inflationary
trajectory with all  fields real and $s=0$ is {\it so simple in the
Jordan frame? }

The answer to this question consists of several parts. (1) The first term
appears directly from our expression for the frame function (\ref{Phi1}).
(2) The kinetic term for scalars at $S=0$ is canonical due to  our choice
of geometric breaking of superconformal symmetry, which does not affect
this important property. (3) The value of the auxiliary field ${\cal
A}_\mu (z, \bar z)$ vanishes for real scalars. (4) The potential in the
Jordan frame (\ref{VJ}) in the $D$-flat direction with cubic
superpotential is
\begin{equation}
V_J= G^{\alpha \bar \beta}W_\alpha  \bar W_{\bar \beta} \,.
\end{equation}
The term $\chi
(H_{u}^{0}H_{d}^{0}+\overline{H_{u}^{0}}\overline{H_{d}^{0}})$ in the frame function signals the deviation from the superconformal theory. This deviation, however, is controllable. Namely, with $W_{\rm Higgs}=-\lambda SH_{u}\cdot H_{d}+{\rho\over 3} S^{3}
$  if we succeed to stabilize the theory at $S=0$,   the only contribution to the potential at $S = 0$ comes from the term
\begin{equation}
V_J|_{S=0}= {\partial W\over \partial S} G^{S \bar S } {\partial   \bar W\over \partial {\bar S}} = \lambda^2
 G^{S \bar S } |H_u\cdot H_d|^2= \lambda^2
 |H_u\cdot H_d|^2 ={\lambda^2\over 16} h^4 \,.
\label{VJ2}
\end{equation}
The metric $G^{S \bar S }=1$ since the field $S$ does not enter in the
superconformal symmetry breaking  $\chi$ term,  and therefore even after
this breaking of superconformal symmetry the $h$-dependent part of the
potential remains quartic: in the $D$-flat direction for real fields it
is equal to ${\lambda^2\over 16} h^4$.

It is easy to compare the supergravity action on inflationary trajectory (\ref{inflation}) with the nonsupersymmetric Jordan frame action (\ref{shap}), which we reproduce here again to simplify the comparison:
\begin{equation}
\mathcal{L}_{J}^{\rm SM}=\sqrt{-g_{J}}\left[ {\frac{1+\xi h^2}2
}R\left( g_{J}\right) -\ft12(\partial_\mu h)^2-{\frac{\lambda
}{4}} h^4 \right] \,. \label{shap2}
\end{equation}

These two actions coincide after the following identification of the parameters:
\begin{equation}
 { \xi ~ \longleftrightarrow ~ -{1\over 6} + {1\over 4} \chi ~ , \qquad \qquad \lambda ~ \longleftrightarrow  ~ {\lambda^2\over 4}} \ .
\label{ident}
\end{equation}
On the left-hand side of each equation in (\ref{ident}) we have parameters of the standard model as in Eq. (\ref{shap2}). On the right-hand side of each equation above we have parameters of the NMSSM inflation model as in Eq. (\ref{inflation}).

After the identification (\ref{ident}), all features of inflation in the NMSSM can be deduced from the results of Ref. \cite{Sha-1} presented in Sec. 3. In particular, the slow-roll parameters are
\begin{eqnarray}
  \label{eq:7a}
  \epsilon &
  \simeq&\frac{64  }{3
   \chi^2h^4}
  \;, \\
  \eta  &
  \simeq& -\frac{16}{3 \chi h^2 } \ .
\end{eqnarray}
Slow roll ends when $\epsilon, \eta \simeq 1$, so the field value at the end of
inflation is
$h_{\mathrm{end}}\simeq 2.2/\sqrt{\chi}$.
The number of e-foldings during the slow roll of the field $h$ from its initial value $h_0$, for $h_0 \gg h_{\mathrm{end}}$, is given by
\begin{equation}
  \label{eq:8a}
  N   \simeq \frac{3}{16}\xi h_0^2
  \;.
\end{equation}
For $N \sim 60$, the amplitude of scalar perturbations of metric corresponds to the COBE normalization for
\begin{equation}
  \label{eq:9a}
  \chi
  \simeq  10^5 {\lambda}
  \;.
\end{equation}
The asymptotic value of the Einstein frame potential $V_E$ at large $h$ is ${\lambda^2\over \chi^2}$, and the Hubble constant during inflation in this model is $H \approx {1\over \sqrt 3} {\lambda\over  \, \chi}$.

To give a particular example, let us take $\lambda = 10^{-2}$. In this case one should have $\chi = 10^{3}$. Inflation ends at $h_{\mathrm{end}} \sim 0.07$. The last 60 e-folds of inflation begin at $h_0 \simeq 0.37$. All observational consequences are the same as in the nonsupersymmetric model   \cite{Sha-1}. In particular, the spectral index is $n_s \sim 0.97$, and the tensor to scalar ratio is $r \approx 0.0033$. These  results are valid for $\chi \gg 1$. They are invariant with respect to the simultaneous rescaling $\lambda \to c\,\lambda$, $\chi \to c\,\chi$ and $h_0\to h_0/\sqrt c$.
For a complete investigation of inflation in this model one would also need to study quantum corrections in supersymmetric theory as it was done for the standard model case in \cite{Sha-1}.


\subsection{Stabilization of the noninflaton directions in the moduli space}

We would like to split all 6 components of the 3 complex scalars $
S,  H_{u}^{0},
H_{d}^{0}$ in (\ref{complex})  into heavy  and light ones. First of all, we  impose a unitary gauge, when one combination of the neutral components of $H_u^0$ and $H_d^0$ is the Goldstone boson and is absent in the unitary gauge. We take  a condition $\alpha_1=\alpha_2$.

We study stabilization of angles $\alpha, \beta, \gamma\equiv  \alpha_1+\alpha_2$ and of the field $s$ using the complete and explicit expressions for the kinetic terms and the potential  in the Einstein frame derived  using the Mathematica code \cite{Kallosh:2004rs} for the \K\, potential in (\ref{K-NMSSM}) and superpotential in (\ref{W-NMSSM}). We present some details of the action in the Jordan frame and the Einstein frame for the real fields $h$ and $s$ in Appendix B.

\subsubsection{Stabilization of angles}

Now we must check the stability of the inflationary solution with respect to the fields $\beta = \pi/4$, $\alpha= 0$, $\gamma\equiv  \alpha_1+\alpha_2=0$, and $s = 0$.
We already checked in \cite{Ferrara:2010yw} that during inflation the $CP$-invariant solution in which $S$, $H_{u}^{0}$ and $H_{d}^{0}$ are
real, is stable with respect to the field $\beta$. The degree of stability is described by the mass squared of the field $\beta$.
During inflation, in the limit $\chi h^2\gg 1$, one has the kinetic term ${2\over \chi} (\partial\beta)^2$ and the second derivative of the potential over $\beta$ is $V_{\beta , \beta} (\beta=\pi/4)={\frac{4(g^2+g^{\prime }{}^2)}{\chi^2  }}$. This means that the effective mass
\begin{equation}
m_\beta^2={\frac{g^2+g^{\prime }{}^2}{\chi }}= {\frac{g^2+g^{\prime }{}^2}{\lambda^2  }} 3 \chi \, H^2 \,,
\end{equation}
is greater than $H^2={1\over 3} {\lambda^2\over  \, \chi^2}$.
In the most natural case $\lambda^2 < 3 \chi \, (g^2+g^{\prime }{}^2)$, one has
$m_\beta^2 \gg H^2$. Thus, there is no slow-roll regime with respect
to the change of $\beta $ during inflation, because the mass squared of
perturbations of the angle $\beta $ is much greater than $H^2={1\over 3} {\lambda^2\over  \, \chi^2}$. During inflation the field $\beta $ rapidly approaches $\pi /4$ and
stays there. For $\lambda^2 \ll  g^2,g^{\prime }{}^2$, the regime with $\beta =\pi/4, \, h_{1}^2=h_2^2=h^2/2$ remains stable even long after inflation, until the soft supersymmetry breaking terms become important and change the final value of $\beta$  \cite{Ferrara:2010yw}.

Now we should study the dependence of the potential on angles $\alpha$ and $ \alpha_1= \alpha_2$ near the inflationary trajectory $s = 0$, $\beta = \pi/4$. The potential  at $s = 0$ does not depend on $\alpha$. Therefore instead  of investigation of excitations of $\alpha$         one should study stability of the potential with respect to the field $s$ for different $\alpha$. For small $s$ and $\lambda \rho <0$, the minimum of the potential with respect to $\alpha$ occurs at  $\alpha = 0$  \cite{Ferrara:2010yw}. As we see later, stability in this direction is achieved by adding the term ${ \zeta\over 3} (S\bar S)^2$ in the K{\"a}hler potential, following the suggestion made in \cite{Lee:2010hj} and ideas developed in \cite{Kitano:2006wz}.

\begin{figure}[!ht]
\centering
\includegraphics[scale=0.2]{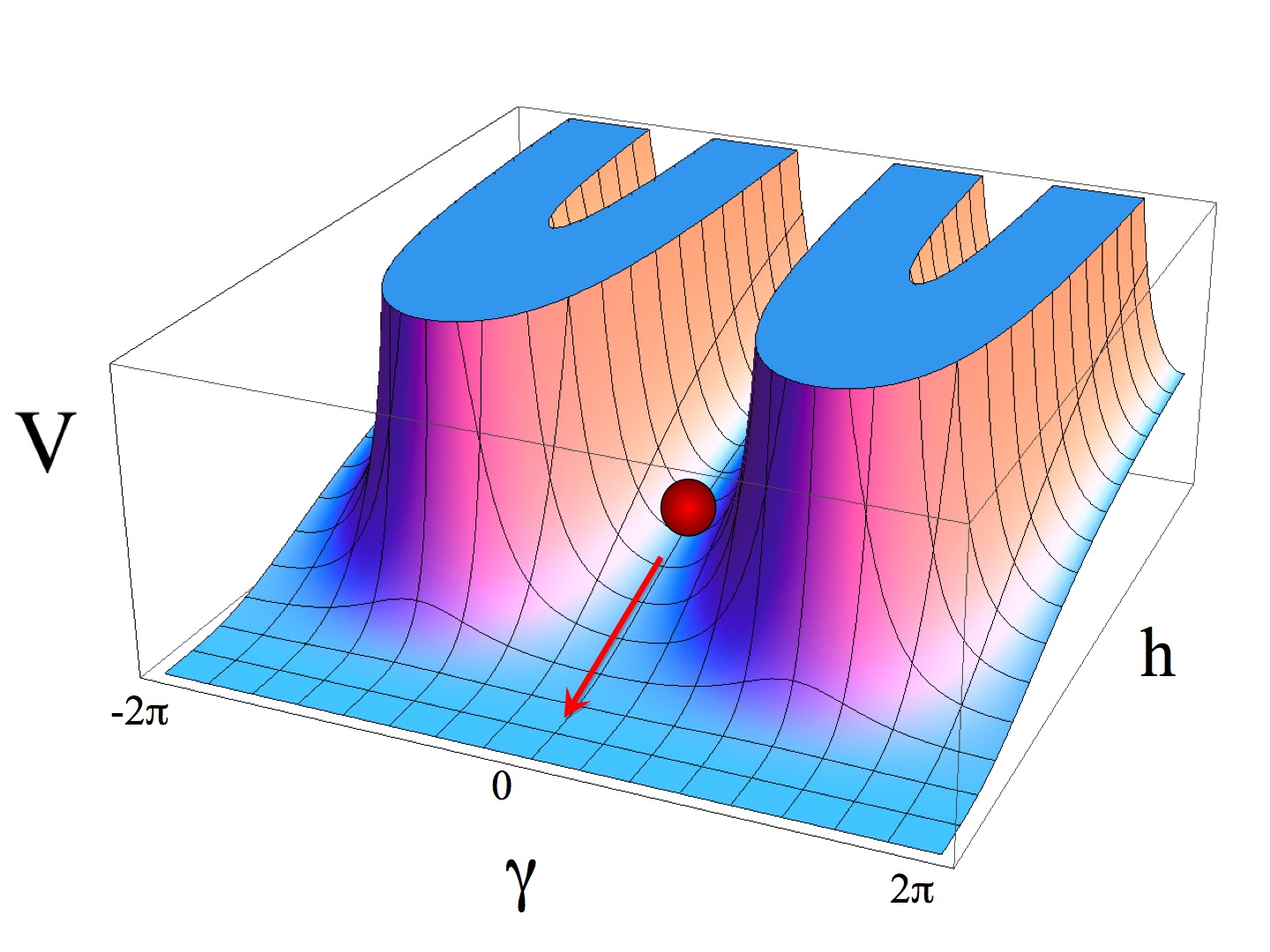}
\caption{Stabilization of the angle $\gamma = \alpha_1+\alpha_2= 0$ near the inflationary trajectory. The infinitely high horseshoe barriers correspond to the singularity of the \K\, geometry. These barriers separate the admissible range of variables from the forbidden part of the landscape (inside the horseshoes), where the argument of the logarithm in the expression for the \K\, potential becomes negative.}
\label{angle}
\end{figure}

As explained above, the  combination $\alpha_1- \alpha_2$ describes a Goldstone boson, which is replaced by a longitudinal component of the vector field. In  the unitary gauge  $\alpha_1- \alpha_2 = 0$.  The remaining combination $\gamma = \alpha_1+ \alpha_2$ corresponds to a scalar field with mass which during inflation is
\be
m^2_\gamma \approx 4 H^2 \,.
\ee
To see it we should analyze the potential along the inflationary direction $s = 0$, $\beta = \pi/4$.
\be
V(h,\gamma) = {9\lambda^2 h^4\over (12-2h^2+3\chi h^2 \cos\gamma)^2}\,,
\ee
where $\gamma = \alpha_1 + \alpha_2$.
At $\chi h^2\gg 1$ $V(h,\gamma) = {\lambda^2 \over \chi^2 \cos^2\gamma}
$.
Therefore its second derivative at $\gamma = 0$ is given by
$
V_{\gamma,\gamma}(\gamma = 0) = {2\lambda^2 \over \chi^2}
$.
The matrix of kinetic terms for the fields $H_i$ in the limit $\chi h^2\gg 1$ at $S=0$ in the unitary gauge  at $\beta = \pi/4$ simplifies to
\bea
{\cal L}_{kin}
& \Rightarrow &  {3\over h^2}(\partial h)^2  +  {3\over 4}(\partial \gamma )^2 \,.
\eea
 Therefore the mass of the canonical field
 \be
m^2_\gamma = {2 \over 3} V_{\gamma,\gamma}(\gamma = 0) =  {4\lambda^2 \over 3\chi^2} =  4V/3 = 4
H^2\,.
\ee
 Since $m^2_\gamma$ is of the same order as $H^2$, during inflation the field $\gamma$ rapidly rolls towards $\gamma = 0$ and stays there. Therefore during inflation we have $\gamma  =0$, or, equivalently, $\alpha_1= \alpha_2 =0$.
During inflation, $\mathbb{Z}_3$ symmetry is  broken by the term $\ft32 \chi (H_u\cdot  H_d + \hc)$ in the
\K\, potential. The potential has a minimum with respect to $\gamma$ only at $\gamma = 0$ (or, more exactly, at $\gamma = 2\pi n$), see Fig. \ref{angle}. Therefore inflation naturally singles out only one of the three possible minima related to each other by $\mathbb{Z}_3$ symmetry. However, long after inflation, when soft supersymmetry breaking terms become important at small $h$ a new strong mechanism of breaking
$\mathbb{Z}_3$ symmetry takes place as we have shown in Sec. \ref{phen}. It originates from the real part of the quadratic holomorphic term $\ft32 \chi (H_u\cdot  H_d + \hc)$ in the
\K\, potential, however, it removes domain walls via the induced soft term in the potential, $V_{\rm soft} \sim \chi  (H_u\cdot  H_d+\hc)$.

\subsubsection{Stabilization of the field $S$}\label{einst}

As we have shown in  \cite{Ferrara:2010yw}, the original version of
inflation in the NMSSM model \cite{Einhorn:2009bh} suffered from the
tachyonic instability with respect to the field $S$. However, one may
circumvent this problem by taking into account interactions of the field
$S$ with superheavy fields that one may add to the model, or simply by
adding a term $-{1\over 3}\zeta (S\bar S)^2$ to the frame function  and
\K\, potential   \cite{Lee:2010hj,Kitano:2006wz}. This term helps to
stabilize the inflationary trajectory in the toy model considered in
\cite{Lee:2010hj}. Here we will check what happens in the  NMSSM model.
An investigation required an extensive use of the Mathematica program
\cite{Kallosh:2004rs}. We will present here only the main results, with
some details given in Appendix \ref{app:Why}.

 During the inflationary regime, the leading behavior of the $F$-term potential in this model for $\xi h^2 \gg 1$ is given by
\begin{equation}
V_{E} \sim {\frac{\lambda^2}{\chi
^2}} -\left(|\lambda \rho |+{\frac{\lambda^2}{6\chi }}(2-3\zeta \chi h^2)\right)
{\frac{4s^2}{\chi^2h^2}}+O(s^{4})\,.
\end{equation}
To find the effective mass of the $s$ field, attention must be paid to
the nonminimal normalization of the field $S=s\rme^{\rmi\alpha }/\sqrt 2$. At
constant $\alpha $, the kinetic term of field $S$ is given by
\begin{equation}
g_{S\overline{S}}\partial S\partial \bar{S}={\frac4{\chi
h^2}}\partial S\partial \bar{S}={\frac2{\chi h^2}}(\partial s)^2\,.
\end{equation}
Here, as we already explained before,  the $\zeta$ correction to the
kinetic term of the $s$-field is always small compared to other terms,
and so we neglected it.  For small $s$, in the vicinity of the
inflationary trajectory, the Lagrangian  of the field $s$ for $\chi h^2
\gg 1$ is
\begin{equation}
\mathcal{L}_{E} \approx   -{\frac 2{\chi h^2}}(\partial s)^2-{
\frac{\lambda^2}{\chi^2}}+\left(|\lambda \rho |+{\frac{\lambda^2}{6\chi }}(2-3\zeta \chi h^2)\right)
{\frac{4s^2}{\chi^2h^2}} \,.
\end{equation}
Therefore the  mass  of the canonical field $s$ is
\begin{equation} m_{{s}}^2\sim 2\left( {\frac{\lambda^2}{6\chi^2}}(3\zeta\chi h^2-2)-{\frac{|\lambda \rho |}{\chi }}\right)\,.
\end{equation}
Thus the condition of stability of the inflationary trajectory at $s = 0$ is
\begin{equation} \zeta >   {2|\lambda\rho|\over \lambda^2 h^2} + {2\over 3\chi h^2}\,.
\end{equation}

\begin{figure}[t!]
\centering
\includegraphics[scale=0.4]{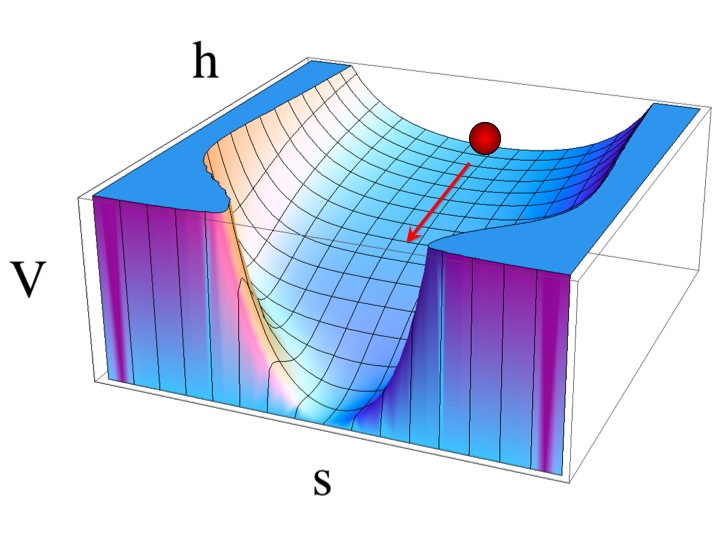}
\caption{Stabilization of the field $s$ near the inflationary trajectory $s = 0$ for $\zeta > {2|\lambda\rho|\over \lambda^2 h^2} + 0.0327$.}
\label{stabil}
\end{figure}
However, this simple analytic form of the bound can be used only  at $\sqrt \chi h \gg 1$, i.e. well before the end of inflation. Meanwhile, the greatest danger of instability occurs at the very end of inflation.
A more accurate condition, which is valid even for  $ \chi h^2 < 1$, is
\begin{equation} \zeta >   {2|\lambda\rho|\over \lambda^2 h^2} + {2(y^2-32)\over 3y^2(y+4)}\,,
\end{equation}
where $y = \chi h^2$.
The function ${2(y^2-32)\over 3y^2(y+4)}$ takes its maximal value $0.0327$ at $y = \chi h^2 \approx 10.9$. This point corresponds to the moment of maximal vulnerability with respect to the tachyonic instability. Therefore the trajectory $s = 0$ remains stable for all $h$ if
\begin{equation} \zeta > {2|\lambda\rho|\over \lambda^2 h^2} + 0.0327\,.
\end{equation}
This result is illustrated by Fig. \ref{stabil}, which shows the potential for $\rho = 0$ and $\zeta = 0.04$.

\begin{figure}[h!]
\centering
\includegraphics[scale=0.27]{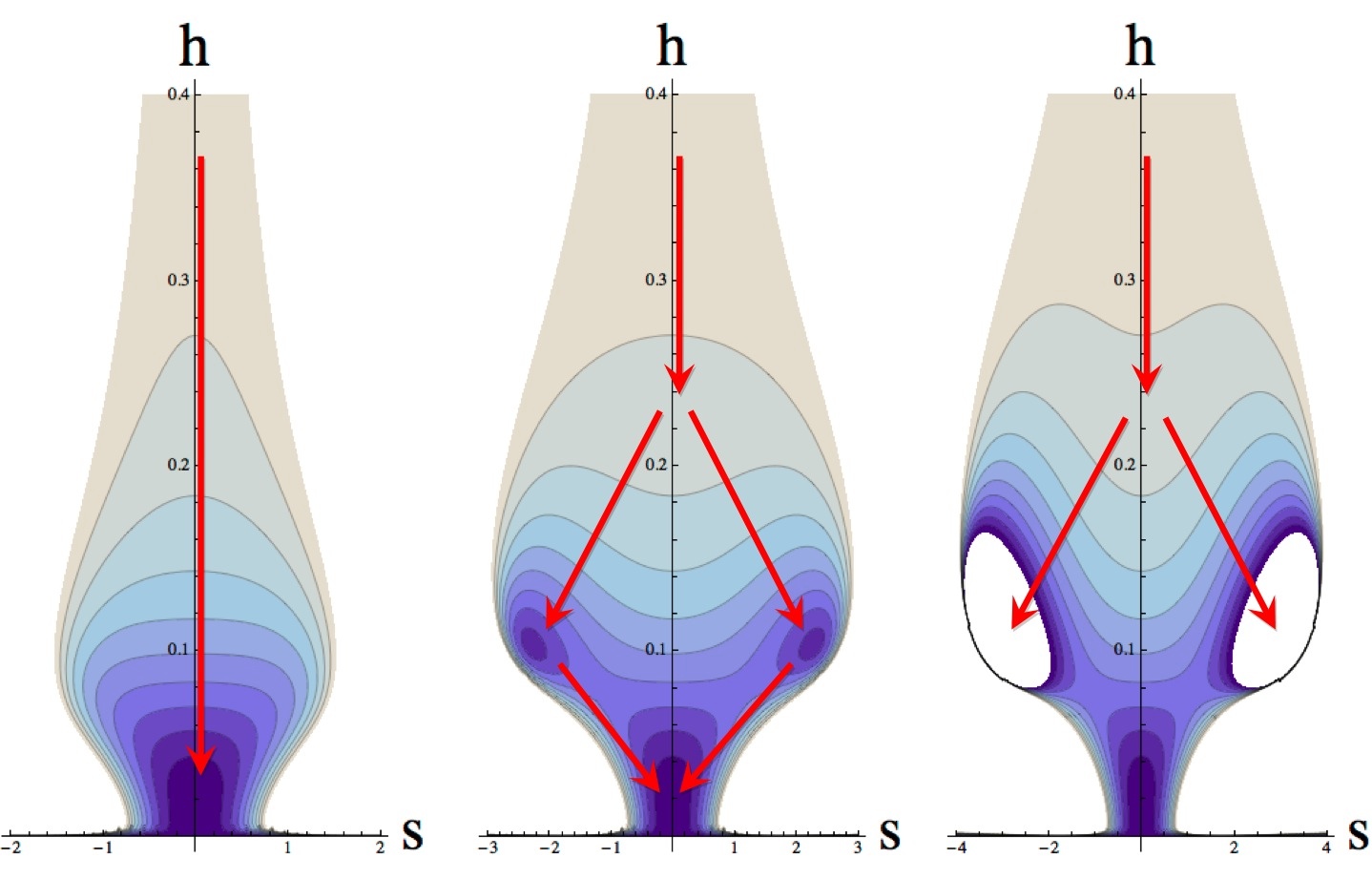}
\caption{Contour plots of the potential for the fields $s$ and $h$ during the last 60 e-folds of inflation for $\rho = 0$,
$\lambda = 10^{-2}$ and $\xi = 5\times 10^2$, for three different values of $\zeta$. The darker parts correspond to smaller positive density,
and the white ovals on the right panel correspond to negative values of the potential. The red arrows show the evolution of the fields during inflation.}
\label{count}
\end{figure}

To illustrate the general situation in a more complete way, we show the
contour plot of the potential of the fields $h$ and $s$ for $\rho = 0$
and various values of the coupling constant $\zeta$ in Fig. \ref{count}.
The first panel corresponds to the potential shown in Fig. \ref{stabil},
with $\zeta = 0.04$. The field $s$ is stabilized during inflation and
after it. For $\zeta$ slightly smaller than $0.0327$,  the tachyonic
instability of the field $s$ at the very end of inflation may force the
field to deviate from the straight path, which results in tachyonic
preheating. This possibility is exotic but not dangerous. Further
decrease of $\zeta$ may result in formation of two local minima of the
potential, as shown in the second panel for  $\zeta = 0.01$. The field
may roll to one of the two metastable minima and be stuck there until it
tunnels to the global minimum at $h,s = 0$. This leads to large
inhomogeneities, as in the old inflation scenario. Finally, the third
panel shows the potential for $\zeta = 0.003$. The field rolls down to
one of the two minima with negative values of the potential, looking like
two white eyes of an alien, and the Universe collapses.

Thus, for $\zeta$ significantly below  ${2|\lambda\rho|\over \lambda^2
h^2} + 0.0327$  one may encounter production of gross inhomogeneities and
even a collapse of the Universe. Fortunately, however, one can have a
successful inflationary scenario if $\zeta$ is greater than
${2|\lambda\rho|\over \lambda^2 h^2} + 0.0327$, and even if $\zeta$ is
slightly below this limit.

\subsection{Higher order corrections}

In Sect. \ref{unitarity} we argued that the unitarity bound discussed in \cite{Burgess:2009ea,Barbon:2009ya,Hertzberg:2010dc} is not expected to pose any problems for the description of inflation in this class of models. However, for the full investigation of these models one should also investigate running of the coupling constants, along the lines of Ref. \cite{Sha-1} where it was done for the standard model case. For our models the quantum corrections have to be studied with account of supersymmetry.

Independently of the issue of quantum field theory-type corrections, one
may wonder how stable are our conclusions with respect to modifications
of various ingredients of these models. According to (\ref{eq:9a}), our
model does require small $\lambda$ and large $\chi$, so that
${\lambda/\chi} \sim 10^{-5}$. This, by itself, does not look much better
than the standard requirement $\lambda \sim 10^{-6}$ in the theory
$\lambda^2\phi^4/16$, as in Eq. (\ref{inflation}). However, the choice of
the large $\chi$ may provide an additional robustness of the model with
respect to higher terms in the expression for the \K\, potential.

Indeed, inflation in our model occurs at $h \sim 1/\sqrt\chi$. Suppose that, in addition to the term  breaking superconformal symmetry, $-\ft12 \chi (H_u\cdot  H_d + \hc)$, there is also a higher order term $c\, (H_u\cdot  H_d + \hc)^2$  in the \K\, potential. For $c = O(1)$, this term remains smaller than the first one for $h < \sqrt \chi$. Therefore if one takes $\chi \gg 1$, then for $h \lesssim \sqrt \chi$ one should not be concerned about higher order terms described above. According to (\ref{eq:8a}), the total number of e-folds in this regime is $N \sim \chi^2$, which is incredibly large for the values of $\chi$ considered in our paper. The total number of e-folds may be even greater if there is some symmetry which protects the original structure of  the \K\, potential at large values of the inflaton field.

\subsection{Gravitino problem and inflation beyond the NMSSM}
The possibility to have an inflationary regime in the NMSSM does not
necessarily mean that the cosmological theory based on this scenario is
fully consistent.  Supergravity is plagued by  the cosmological moduli
problem and by the gravitino problem. Inflation helps to solve the
gravitino problem, but only if the reheating temperature $T_{r}$ after
inflation is sufficiently small. The bounds on $T_{r}$ depend on the
gravitino mass and other parameters, but typically it should be smaller
than $10^{8}$ GeV, see
\cite{Ellis:1982yb,Khlopov:1984pf,Moroi:1995fs,Ferrantelli:2010as,Kawasaki:2008qe,Hasenkamp:2010if}
for a more detailed discussion of this issue.

One way to avoid this problem is to assume that the energy scale of
inflation is very low, which leads to a small reheating temperature.
However, inflation in the NMSSM occurs at the energy density
${\lambda^{2}\over \chi^{2}} \sim 10^{-10}$, in Planck units. If
reheating happens instantly, this energy is converted to thermal energy
$T_{r}^{4} \sim 10^{{-10}}$. This gives an estimate $T_{r} \sim
10^{{15}}$ GeV.

One may have a much smaller reheating temperature if the inflaton field
extremely weakly couples to matter, which leads to a delay in
thermalization. During this delay, the energy of the inflaton field
decreases, and the reheating temperature becomes smaller. For example,
one may consider inflationary models where the inflaton belongs to a hidden
sector, and its decay to observable particles is suppressed by the small
gravitational coupling. But the Higgs fields belong to the observable
sector and they couple to matter quite strongly. An investigation of
reheating in the Higgs inflation \cite{Bezrukov:2008ut} suggested that
the reheating temperature is about $10^{13}$ GeV. A more detailed
investigation performed in \cite{GarciaBellido:2008ab} demonstrated that
the process of reheating in this theory is quite complex, being a
combination of the perturbative reheating \cite{reheating}, parametric
resonance \cite{KLS94}, instant preheating \cite{InstantFKL98}, and
tachyonic preheating \cite{Tachyonic2001}. The authors argued that the
full investigation of this complicated process should be done by lattice
simulations \cite{Khlebnikov:1996mc}. However, there is no obvious reason
to expect that this investigation will yield the reheating temperature 5
orders of magnitude smaller than the estimate $T_{r} \sim 10^{{13}}$ GeV made in
\cite{Bezrukov:2008ut}.

There are several possible ways to address this problem, even if the
future investigation confirms that $T_r \gg 10^8$ GeV. First of all, the
gravitino problem disappears if the gravitino mass is below keV, or if it
is several orders of magnitude above the TeV scale; see e.g.
\cite{Ferrantelli:2010as,Kawasaki:2008qe,Hasenkamp:2010if} for a recent
discussion and more precise bounds on the gravitino mass. Both of these
possibilities are realistic. For example, in the model of conformal gauge
mediation one can have gravitino with mass below 10 eV \cite{Ibe:2008si}.
Superheavy gravitinos have been discussed in
\cite{DeWolfe:2002nn,Arkani-Hamed:2004fb,Blumenhagen:2009gk}.

Another solution is to have a second stage of inflation after the NMSSM
inflation. This is a realistic possibility since the energy scale of the
NMSSM inflation is very high, so it is quite possible to have a second
stage of inflation at a much smaller energy scale after the NMSSM
inflation. If this stage is short, as in the thermal inflation scenario
\cite{thermal}, then it may solve the gravitino problem, and all
observational predictions on the NMSSM inflation will remain intact. On
the other hand, if the second stage of inflation is sufficiently  long,
then it will determine all properties of the observable part of the
Universe. In this respect, it is quite encouraging that one can develop a
large class of new models of chaotic inflation based on the ideas
discussed in this paper, but without necessarily identifying the inflaton
field with the Higgs field of the standard model  \cite{Kallosh:2010ug}.

\section{Conclusion}

Supergravity phenomenology was mostly developed in the Einstein frame
where there is no  scalar-curvature coupling. In this paper we propose a
superconformal approach to supergravity phenomenology and cosmology. One
can start with the $SU(2,2|1)$ superconformal theory of  chiral  and
vector multiplets interacting with supergravity Weyl multiplet.  This
theory contains a conformal compensator, which can be gauged away, giving
rise to the Planck mass.  In this paper we identified a special class of
supergravity models:  If chiral and vector multiplets of the
superconformal theory are decoupled from the conformal compensator, the
part of the action describing matter fields in the Jordan frame remains
superconformal invariant. This action is unusually simple: kinetic terms
are canonical, supergravity potential coincides with the global theory
potential, and scalars are conformally coupled to gravity. The potential
is quartic, the theory has no mass terms, no nonrenormalizable terms, and
no cosmological constant.

Theories of this type may form a convenient starting point for
constructing phenomenological models. In such models, one may attribute
smallness of all mass parameters to the effects of breaking of the
superconformal symmetry, which can be achieved e.g. due to gravitational
effects. In particular, these theories may provide a natural supergravity
embedding for the NMSSM. Superconformal symmetry breaking is introduced
by the real part of the holomorphic quadratic nonminimal scalar-curvature
coupling, by terms designed to stabilize some fields at the origin of
moduli space, and by interactions with a hidden sector. This approach to
supergravity phenomenology from the underlying superconformal theory
allows one to address the $\mu$ problem and the domain wall problem, and
to obtain an inflationary regime in the NMSSM. Efficient reheating after
inflation in the NMSSM may lead to the cosmological gravitino problem.
This problem can be solved if one considers models with superlight or
superheavy gravitino, or if one postulates a secondary stage of inflation
after the NMSSM inflation. Fortunately, the general methods developed
during the investigation of the canonical superconformal supergravity and
inflation in the NMSSM can be used for construction of a new broad class
of models of chaotic inflation in supergravity with a functional freedom
of choice of the inflaton potential \cite{Kallosh:2010ug}.

\section*{Acknowledgments}

We are grateful to S.~A.~Abel, S.~Dimopoulos, M.~Einhorn,  U.~Ellwanger,
J.~Garcia-Bellido, C.~Germani, P.~Graham,  A.~Guth, S.~Kachru, B. Kyae,
H.~M.~Lee, P. Nilles, K. Olive, F.~Quevedo, S.~Shenker, L.~Susskind and
A.~Westphal
 for useful discussions. The work of R.K. and A.L. is supported by
NSF Grant No. 0756174. The work of S.F. is supported by ERC Advanced
Grant No. 226455, \textit{Supersymmetry, Quantum Gravity and Gauge
Fields} (\textit{Superfields}), and in part by INFN, sez. L.N.F. The work
of A.M. is supported by  INFN, as a visitor to
 Stanford University,
Stanford, CA, USA. The work of A.V.P. is supported in part by the FWO -
Vlaanderen, Project No. G.0235.05, and in part by the Federal Office for
Scientific, Technical and Cultural Affairs through the
``Interuniversity Attraction Poles Programme -- Belgian Science
Policy''  P6/11-P.


\appendix

\section{Complete CSS action}
 \label{app:completeAct}

We present here the full action corresponding to the 4 assumptions given
in Sec. \ref{scmodes}. We will eliminate the scalar auxiliary fields of
supergravity, but leave the auxiliary vector $A_\mu $ as an independent
field. We will make use of the gauge conditions mentioned in Sec.
\ref{gfix}.

Because of the separation of $X^0$ from the other fields, the action can
be split and we can write
\begin{eqnarray}
  S&=&\int \rmd^4x\,\left\{ [-|X^0|^2+|X^\alpha|^2]_D + [\ft13d_{\alpha \beta \gamma }X^\alpha X^\beta X^\gamma ]_F + \left[ f_{AB} \bar \lambda ^A P_L \lambda^B\right] _F\right\} \nonumber\\
&=&\int \rmd^4x\,\left[  {\cal L}_{\rm SG}+{\cal L}_{\rm conf}\right] \,,
\label{SSSG+conf}
\end{eqnarray}
where $S_{\rm SG}$ is the action of pure supergravity, which is produced
from the $|X^0|^2$-term, and $S_{\rm conf}$ is the conformal action of
all the other physical fields. The former is given by
\begin{equation}
  {\cal L}_{\rm SG}=\,e\,\ft12M_P^2\left[ R(\omega (e,\psi )) -\bar \psi _\mu \gamma ^{\mu \nu \rho }
  \left( \partial _\nu +\ft14 \omega _\nu {}^{ab}(e,\psi )\gamma _{ab}\right) \psi _\rho+6A^\mu A_\mu  \right]\,,
 \label{resultSconfN0pure}
\end{equation}
where we already replaced $X^0$ by its gauge-fixed value $\sqrt 3 \, M_P$.
The conformal $D$-terms are
\begin{eqnarray}  [|X^\alpha|^2 ]_D=&e\,\delta _{\alpha \bar \beta }&\left\{
 -{\cal D} _\mu X^\alpha  {\cal D}^\mu \bar X^{\bar \beta }-\frac12\bar \Omega^\alpha
  \slashed{\cal D }\Omega^{\bar \beta }-\frac12\bar \Omega^{\bar \beta }
  \slashed{\cal D }\Omega^\alpha +F^\alpha  \bar F^{\bar\beta  }\right.
\nonumber\\
&&- \rmi \bar X^{\bar \beta  }k_A{}^\alpha  D^A -\sqrt{2}  \bar \lambda ^A
  \left( \Omega ^\alpha  k_A{}^{\bar \beta }
   +\Omega ^{\bar \beta }k_A{}^\alpha \right)
\nonumber\\
&&+\left[ \frac1{2\sqrt{2}}\bar \psi \cdot \gamma
\left(F^\alpha \Omega^{\bar \beta }-\slashed{\cal D}\bar X^{\bar \beta }\Omega^\alpha
+X^{\bar \beta }P_R\lambda ^A k_A{}^\alpha \right)-\frac{1}{6\sqrt{2}} \bar X^{\bar \beta } \bar \Omega^\alpha
\gamma ^{\mu \nu } R'_{\mu
\nu }(Q) \right.
\nonumber\\
&&\left.+\frac{1}{8}\rmi\varepsilon ^{\mu \nu \rho \sigma }\bar \psi _\mu \gamma _\nu \psi _\rho
\left( \bar X^{\bar \beta }{\cal D}_\sigma X^\alpha +\frac{1}{2}\bar \Omega^\alpha \gamma _\sigma \Omega^{\bar \beta }
+\frac{1}{\sqrt{2}}\bar X^{\bar \beta }\bar \psi _\sigma \Omega^\alpha \right)
+\hc\right]\nonumber\\
&&\left.+\frac16X^\alpha \bar X^{\bar  \beta }\left(-R(\omega(e,\psi ) )
+\ft12\bar \psi _\mu \gamma ^{\mu \nu \rho } R'_{\nu \rho }(Q)
\right) \right\} \,.
\end{eqnarray}
The fermions are chiral:
\begin{equation}
  P_L\Omega ^\alpha =\Omega ^\alpha \,,\qquad P_R\Omega ^{\bar \beta }=\Omega ^{\bar \beta }\,.
 \label{chiralfermions}
\end{equation}
The covariant curvature and covariant derivatives in this equation are
\begin{eqnarray}
 {\cal D}_\mu X^\alpha  &=& \left( \partial _\mu  - \rmi A_\mu\right)  X^\alpha
  -\frac{1}{\sqrt{2}}\bar \psi _\mu \Omega^\alpha -A_\mu ^Ak_A{}^\alpha \,, \nonumber\\
  {\cal D}_\mu  \Omega^\alpha  &=&\left( \partial _\mu+\frac14\omega _\mu {}^{ab}(e,\psi )\gamma _{ab}
  +\frac12\rmi A_\mu\right)
 \Omega^\alpha  -\frac1{\sqrt{2}} \left(\slashed{\cal D} X^\alpha  +
F^\alpha \right)\psi _\mu  -\sqrt{2} X^\alpha \phi _\mu 
-A_\mu ^A(m_A)^\alpha {}_\beta\Omega^\beta  \,,\nonumber\\
R'_{\mu \nu }(Q)&=&2\left( \partial _{[\mu }-\ft32\rmi
A_{[\mu }\gamma _*+\ft14\omega _{[\mu }{}^{ab}(e,\psi )\gamma _{ab}\right) \psi
_{\nu ]}\,.
 \label{covderchiraltot}
\end{eqnarray}
The spin connection $\omega _\mu {}^{ab}$ in this equation contains
$\psi$ torsion. The field $\phi _\mu $ is the composite gauge field of
special supersymmetry:
\begin{equation}
  \phi _\mu =-\ft12 \gamma ^\nu  R'_{\mu \nu }(Q)+\ft{1}{12}\gamma _\mu \gamma
  ^{ab}R'_{ab}(Q)\,.
 \label{solphimu}
\end{equation}

The superpotential term is
\begin{eqnarray}
[\ft13d_{\alpha \beta \gamma }X^\alpha X^\beta X^\gamma]_Fe^{-1}&=& d_{\alpha \beta \gamma }X^\alpha X^\beta
F^\gamma  -d_{\alpha \beta \gamma }X^\alpha\bar \Omega^\beta \Omega ^\gamma
 +d_{\alpha \beta \gamma }X^\alpha X^\beta\bar \psi\cdot
\gamma  \Omega^\gamma \nonumber\\
&&+\ft16 d_{\alpha \beta \gamma }X^\alpha X^\beta X^\gamma\bar \psi _{\mu }P_R \gamma ^{\mu \nu
}\psi _{\nu }+ \hc\,.
 \label{superpotd}
\end{eqnarray}
The superconformal-invariant kinetic terms for the gauge multiplets are
\begin{eqnarray}
\lefteqn{\left[f_{AB }\bar \lambda^A P_L  \lambda ^B\right]
_Fe^{-1}=-\frac14f_{AB}\left[2\bar \lambda ^AP_L\slashed{\cal D}\lambda
^B+\widehat F_{\mu \nu }^{-A }\widehat F^{\mu \nu \,-B } - D^A
D^B\right.}\nonumber\\&&\left.+\frac{1}{4} \bar \psi\cdot \gamma
P_L\left( \frac12\gamma ^{\mu \nu } \widehat F_{\mu \nu }^{-A }-\rmi D^A
\right)\lambda^B-\frac{1}{8}\bar \psi _\mu \gamma ^{\mu \nu }P_R\psi _\nu
\bar \lambda ^AP_L\lambda ^B\right]
+ \hc \,.
\label{fullaction}
\end{eqnarray}
Here
\begin{eqnarray}
  \widehat{F}_{\mu \nu }{}^A&=& 2\partial _{[\mu }A_{\nu
  ]}{}^A  +g f_{BC}{}^AA_\mu{}^B
A_\nu{}^C+\bar \psi _{[\mu }  \gamma_{\nu ]}\lambda ^A \,,\nonumber\\
  {\cal D}_\mu \lambda ^A&\equiv& \left( \partial_\mu +\ft14\omega _\mu {}^{ab}(e,\psi )\gamma _{ab}-\ft{3}{2}\rmi\gamma _* A_\mu\right) \lambda ^A  +\lambda ^C A_\mu{}^B
 f_{BC}{}^A-\left[\ft{1}{4} \gamma^{ab}\widehat{F}_{ab}{}^A
+\ft{1}{2}\rmi \gamma_*
D^A\right]\psi _\mu \,.
 \label{sconformalFDlambda}
\end{eqnarray}

Many cancellations occur in terms with gravitinos when the various
covariantizations are written in detail, and the torsion terms are
extracted from the spin connection. The supergravity action is then
\begin{eqnarray}
  {\cal L}_{\rm SG}&=&e\,\ft12M_P^2\left[ R(\omega (e)) -\bar \psi _\mu \gamma ^{\mu \nu \rho }
  \left( \partial _\nu +\ft14 \omega _\nu {}^{ab}(e )\gamma _{ab}\right) \psi _\rho+6A^\mu A_\mu +{\cal L}_{\rm SG,torsion}
  \right]\,,\nonumber\\
  &&{\cal L}_{\rm SG,torsion}=-\ft{1}{16}\left[
(\bar{\psi}^\rho\gamma^\mu\psi^\nu) ( \bar{\psi}_\rho\gamma_\mu\psi_\nu
+2 \bar{\psi}_\rho\gamma_\nu\psi_\mu) - 4 (\bar{\psi}_\mu
\gamma\cdot\psi)(\bar{\psi}^\mu \gamma\cdot\psi)\right]\,.
 \label{SSGwithtorsion}
\end{eqnarray}
We now choose the physical scalars and fermions
\begin{equation}
  z^\alpha =X^\alpha \,,\qquad \chi ^\alpha =\Omega ^\alpha \,.
 \label{physicalzchi}
\end{equation}
After elimination of the auxiliary fields $F^\alpha $ and $D^A$,
the conformal part of the action becomes
\begin{eqnarray}
 e^{-1}{\cal L}_{\rm conf}  = &\delta _{\alpha \bar \beta }&\left\{
 -D _\mu z^\alpha  D^\mu \bar z^{\bar \beta }-\ft12\bar \chi^\alpha
  {\slashed{D}}\chi^{\bar \beta }-\ft12\bar \chi^{\bar \beta }
  {\slashed{D }}\chi^\alpha -F^\alpha \bar F^{\bar \beta }\right.
\nonumber\\
&&+\ft16 z^\alpha \bar z^{\bar  \beta }
\left[ -R(\omega (e)) + \bar \psi_\mu R^\mu+ \edet^{-1}
\partial _\mu (\edet \bar \psi \cdot \gamma \psi ^\mu)-{\cal L}_{\rm SG,torsion}\right]
\nonumber\\
&&+\left[ \ft{1}{8}\rmi\varepsilon ^{\mu \nu \rho \sigma }\bar \psi _\mu \gamma _\nu \psi _\rho
 \bar z^{\bar \beta }D_\sigma z^\alpha
 \right.\nonumber\\
&& +\frac{1}{\sqrt{2}} \bar \psi_\mu \slashed{D}
z^{\bar \beta } \gamma^\mu \chi^\alpha
-\frac2{3\sqrt{2}}\bar z^{\bar \beta }\bar \chi^\alpha    \gamma ^{\mu\nu} \left(
\partial _\mu  +\ft14 \omega _\mu  {}^{ab}(e,\psi )\gamma _{ab}
 -\ft32\rmi A_\mu  \gamfive\right)\psi _\nu
 \nonumber\\
&&\left.+\left( -\sqrt{2}   \bar \chi^{\bar \beta }
-\ft12 \bar  \psi  \cdot \gamma P_L  z^{\bar \beta }\right) \lambda^A(m_A)^\alpha {}_\gamma  z^\gamma
+\hc\right]\nonumber\\
&&\left.
+\ft{1}{16}\rmi\,\edet^{-1}\varepsilon^{\mu\nu\rho\sigma}\bar
 \psi _\mu \gamma_\nu \psi _\rho\,\bar \chi ^{\bar \beta }\gamma _\sigma \chi ^\alpha
-\ft12  \bar \psi _\mu \chi ^{\bar \beta }\,\bar \psi _\mu \chi ^\alpha
 \right\}  \nonumber\\
&+(\Re f_{A B})&\left\{ -\ft14 F_{\mu \nu }^A F^{\mu \nu
\,B } -\ft12 \bar \lambda ^A \slashed{D} \lambda ^B-\ft12 D^AD^B\right.\nonumber\\
&&\left.+\ft18\bar \psi _\mu \gamma ^{ab}\left( F_{ab}^A+ \widehat F_{ab}^A \right)
\gamma ^\mu \lambda ^B
+\ft{1}{32}\rmi\,\edet^{-1}\varepsilon^{\mu\nu\rho\sigma}\bar
 \psi _\mu \gamma_\nu \psi _\rho \bar \lambda ^A\gamfive\gamma _\sigma \lambda ^B
\right\}\nonumber\\
&+\left\{\phantom{{}^\alpha }d_{\alpha \beta \gamma }\right.&\left.\left[ -z^\alpha\bar \chi^\beta \chi ^\gamma
 +z^\alpha z^\beta\bar \psi\cdot
\gamma  \chi^\gamma +\ft16z^\alpha z^\beta z^\gamma\bar \psi _{\mu }P_R \gamma ^{\mu \nu
}\psi _{\nu }\right] + \hc\right\} \,.
 \label{confactdelta}
\end{eqnarray}
The covariant derivatives $D_\mu $ have no torsion in the spin
connection, neither supersymmetric covariantization. The same is true for
$F_{ab}$ and $R^\mu $:
\begin{eqnarray}
D_\mu z^\alpha  & = & \left( \partial _\mu  -\rmi A_\mu\right) z^\alpha -A_\mu
^A m_A {}^\alpha{}_\beta z^\beta
  \,, \nonumber\\
D_\mu \chi^\alpha  &=&\left(
\partial _\mu +\ft14 \omega _\mu {}^{ab}(e)\gamma _{ab}
 +\ft 12\rmi A_\mu\right)\chi ^\alpha -A_\mu ^A m _A {}^\alpha{}_\beta  \,\chi  ^\beta \,,
   \nonumber\\
D_\mu \lambda ^A &=&\left( \partial _\mu+\ft14 \omega
_\mu {}^{ab}(e)\gamma _{ab} -\ft 32\rmi A_\mu \gamfive\right)
\lambda ^A  -A_\mu ^C\lambda ^B f_{BC}{}^A \,,\nonumber\\
F_{\mu \nu }{}^A&=&2\partial _{[\mu }A_{\nu
  ]}{}^A  +g f_{BC}{}^AA_\mu{}^B
A_\nu{}^C \,,\nonumber\\
  R^\mu &\equiv& \gamma ^{\mu \rho \sigma }\left(
\partial _{\rho }  +\ft14 \omega _{\rho } {}^{ab}(e)\gamma _{ab}
 -\ft32\rmi A_\rho \gamfive\right)\psi _\sigma \,.
 \label{defRmu}
\end{eqnarray}

The auxiliary fields $A_\mu $ in (\ref{confactdelta}) are to be
considered as independent fields, which should still be solved for by
their field equations. The latter will mix the supergravity part and the
superconformal part of the action. The fields $F^\alpha $ and
$D^A$ on the other hand are to be considered as their expressions in
terms of the other fields:
\begin{eqnarray}
 \bar F^{\bar \beta } & = & - \delta^{\alpha \bar \beta }d_{\alpha \beta \gamma }z^\beta z^\gamma \,, \nonumber\\
 D^A & = & (\Re f)^{-1\,AB}{\cal P}_A= (\Re f)^{-1\,AB}\rmi\delta _{\alpha \bar \beta }z^{\bar \beta }(m_A)^\alpha {}_\gamma  z^\gamma\,.
 \label{valueFD}
\end{eqnarray}
This does not mix the supergravity and the superconformal part. Thus,
Eqs. (\ref{SSGwithtorsion})-(\ref{valueFD}) provide the generalization of
Eqs. (\ref{L0}) and (\ref{VJ0}) when all fermions and vectors are
included.

\section{Why is the supergravity potential in CSS Jordan frame the same as in global SUSY?}
 \label{app:Why}

Starting from the superconformal theory potential in  (\ref{confactpot})
we have already derived the potential of the CSS in  (\ref{VJ0}). The main reason from that point of view is that
the modifications to the global SUSY potential originate from the
compensating multiplet, containing the scalar $(y\bar y)^3=\rme^{{\cal
K}}$, and the auxiliary field $F^0$ producing the term $-3|W|^2$. This compensating multiplet has been decoupled in CSS.  The
fact that for the CSS models the supergravity potential is the same as in globally
supersymmetric models with canonical kinetic terms is somewhat
surprising, from the point of view of the complicated Einstein frame
$F$-term potential in  generic supergravity theory
\begin{equation}
V_{E}=  \rme^{\mathcal{K}}\left(
\nabla _\alpha Wg^{\alpha \bar\beta }\nabla _{\bar\beta }\overline{W} -3W\overline{W}
\right) \,,
\end{equation}
where
\begin{equation}
\nabla_{\alpha }W\equiv W_{\alpha }+ {K}_{\alpha} W \, .  \label{8}
\end{equation}
It is therefore instructive to see directly how the cancellation of
various terms in the $F$-term potential  takes place, leading to a simple
CSS Jordan frame potential.

 We define the Jordan frame for the CSS via the frame function $\Phi \left( z,\overline{z}\right)= -3\, \Omega^2$ related to the \K\, potential $
\mathcal{K}\left( z,\overline{z}\right) =-3\log \Omega^2
$.
The metric in the Einstein frame is related to the metric in the  Jordan frame  as $g_{\mu\nu}^E= \Omega^2 g_{\mu\nu}^J$ and $\sqrt {g^E} = \Omega^4 \sqrt {g^J}$.
The $F$-term potential in the Jordan frame specified by the frame function (\ref{frame}) is related to the Einstein frame potential as
\be V_{J} = \Omega^4 V_E = \Omega^4 \rme^{\mathcal{K}}\left(
\nabla _\alpha Wg^{\alpha \bar\beta }\nabla _{\bar\beta }\overline{W} -3W\overline{W}
\right)\,.
\ee
 We take into account that in CSS
 \begin{eqnarray}
\Omega^4 \rme^{\mathcal{K}}=\Omega^{-2} \,,
\label{frame}\end{eqnarray}
 which means that
 \begin{equation}
V_{J} =\Omega^{-2}\left(
\nabla _\alpha Wg^{\alpha \bar\beta }\nabla _{\bar\beta }\overline{W} -3W\overline{W}
\right)\,.  \label{7}
\end{equation}
 In these models we  have  the following \K\, potential and generic cubic superpotential:
\begin{eqnarray}
\mathcal{K}\left( z,\overline{z}\right)
&=&-3\log \left( 1-\ft{1}{3}\delta _{\alpha \overline{\beta }}z^{\alpha }%
\overline{z}^{\overline{\beta }}\right)\,, \qquad W(z) =\ft{1}{3}d_{\alpha \beta \gamma }z^{\alpha }z^{\beta }z^{\gamma }\,,  \label{2-bis}
\end{eqnarray}
and
\be
\Omega^2= 1-\ft{1}{3}\delta _{\alpha \overline{\beta }}z^{\alpha }%
\overline{z}^{\overline{\beta }}\,.
\ee
It follows that the \K\, geometry with $g_{\alpha \overline{\beta }}g^{\alpha \overline{\gamma }}=\delta _{
\overline{\beta }}^{\overline{\gamma }}$ has the following properties:
\begin{equation}
\mathcal{K}_{\alpha } =\rme^{(1/3)\mathcal{K}}\delta _{\alpha
\overline{\beta }}\overline{z}^{\overline{\beta }}\,, \qquad
g^{\alpha \overline{\beta }} =\rme^{(-1/3)\mathcal{K}}\left( \delta
^{\alpha \overline{\beta }}-\ft{1}{3}z^{\alpha }\overline{z}^{\overline{%
\beta }}\right) = \Omega^2 \left( \delta
^{\alpha \overline{\beta }}-\ft{1}{3}z^{\alpha }\overline{z}^{\overline{%
\beta }}\right)\,, \label{5}
\end{equation}
 so that
\begin{equation}
V_{J} =
\Omega^{-2} g^{\alpha
\overline{\beta }}\left( W_{\alpha }+W\mathcal{K}_{\alpha
}\right) \left( \overline{W}_{\overline{\beta }}+%
\overline{W}\mathcal{K}_{\overline{\beta }}\right) - 3 \Omega^{-2} \left| W\right| ^{2}\,.
\end{equation}
With an account of the CSS \K\, geometry properties we may rewrite the potential as  follows:
\begin{equation}
V_{J}=\delta ^{\alpha \overline{\beta }}W_{\alpha }\overline{W}%
_{\overline{\beta }}-3\left| W\right| ^{2}+\overline{W}W_{\alpha
}z^{\alpha }+W\overline{W}_{\overline{\alpha }}\overline{%
z}^{\overline{\alpha }}-\ft{1}{3}\left| W_{\alpha }z^{\alpha
}\right| ^{2}\,.  \label{9}
\end{equation}
Note that all $\Omega^{-2}$ factors  in (\ref{9}) have canceled. It
remains to take into account that the CSS superpotential $W$ is
homogeneous of the third degree  in $z^{\alpha }$'s, and it follows that
\be
W_{\alpha }z^{\alpha }=3W \,,\qquad  \overline{W}_{\overline{%
\alpha }}\overline{z}^{\overline{\alpha }}=3\overline{W}\,.
\label{10}
\ee
This allows one to bring the $F$-term potential to the final form
\be
V_{J}=\delta ^{\alpha \overline{\beta }}W_{\alpha }\overline{W}%
_{\overline{\beta }}\,,  \label{11}
\ee
where it is clear that it  coincides with the global supersymmetric $F$-term potential.

\section{The moduli space geometry in CSS models with symmetry breaking $%
\protect\chi$ terms}
 \label{app:G}
 In CSS models the moduli space geometry is flat as shown in (\ref{flat}). When superconformal symmetry is broken by the $\chi $-terms of the form given in (\ref{J}), the Jordan frame potential depends on $%
G^{\alpha \bar{\beta}}$ according to Eq. (\ref{VJ}). Here we study the nonflat geometry for the models in (\ref{J}) and, in particular, we compute $G^{\alpha \bar{\beta}}$.
 We start from Eq. (\ref{J}) which we repeat here for convenience
\begin{equation}
\mathcal{N}\left( X,\bar{X}\right) =-\left| X^{0}\right| ^{2}+\left|
X^{\alpha }\right| ^{2}-\chi \left( a_{\alpha \beta }\frac{X^{\alpha
}X^{\beta }\bar{X}^{\bar{0}}}{X^{0}}+\bar{a}_{\bar{%
\alpha }\bar{\beta }}\frac{\bar{X}^{\bar{\alpha }}\bar{X}%
^{\bar{\beta }}X^{0}}{\bar{X}^{\bar{0}}}\right)\,.
\end{equation}
The metric
\begin{equation}
G_{I\bar{J}}\equiv \frac{\partial ^{2}\mathcal{N}}{\partial
X^{I}\partial \bar{X}^{\bar{J}}}\,,
\end{equation}
can be computed to read
\begin{eqnarray}
G_{0\bar{0}} &=&-1+\chi \left[ a_{\alpha \beta }\frac{X^{\alpha
}X^{\beta }}{\left( X^{0}\right) ^{2}}+\bar{a}_{\bar{\alpha }%
\bar{\beta }}\frac{\bar{X}^{\bar{\alpha }}\bar{X}^{%
\bar{\beta }}}{\left( \bar{X}^{\bar{0}}\right) ^{2}}\right]\,,
\nonumber\\
G_{0\bar{\beta }} &=&-2\chi \bar{a}_{\bar{\beta }\bar{%
\gamma }}\frac{\bar{X}^{\bar{\gamma }}}{\bar{X}^{\bar{0}}%
}\,,\nonumber\\
G_{\alpha \bar{0}} &=&-2\chi a_{\alpha \gamma }\frac{X^{\gamma }}{X^{0}}%
\,,\nonumber\\
G_{\alpha \bar{\beta }} &=&\delta _{\alpha \bar{\beta }}\,.
\end{eqnarray}
The components of the inverse metric $G^{I\bar{J}}$ (such that $G^{I%
\bar{J}}G_{I\bar{K}}=\delta _{\bar{K}}^{\bar{J}}$) can be computed to
read
\begin{eqnarray}
G^{0\bar{0}} &=&-\frac{\left( X^{0}\bar{X}^{\bar{0}}\right)
^{2}}{\left[
\begin{array}{l}
\left( X^{0}\bar{X}^{\bar{0}}\right) ^{2}-\chi a_{\gamma \eta
}X^{\gamma }X^{\eta }\left( \bar{X}^{\bar{0}}\right) ^{2}-\chi
\bar{a}_{\bar{\gamma }\bar{\eta }}\bar{X}^{\bar{%
\gamma }}\bar{X}^{\bar{\eta }}\left( X^{0}\right) ^{2} \\
+4\chi ^{2}X^{0}\bar{X}^{\bar{0}}\delta ^{\gamma \bar{\eta }%
}a_{\gamma \theta }\bar{a}_{\bar{\eta }\bar{\rho }}X^{\theta }%
\bar{X}^{\bar{\rho }}
\end{array}
\right] }\,,\nonumber\\
&&  \notag \\
G^{0\bar{\beta }} &=&-\frac{2\chi \delta ^{\lambda \bar{\beta }%
}a_{\lambda \xi }X^{\xi }\left( \bar{X}^{\bar{0}}\right) ^{2}X^{0}%
}{\left[
\begin{array}{l}
\left( X^{0}\bar{X}^{\bar{0}}\right) ^{2}-\chi a_{\gamma \eta
}X^{\gamma }X^{\eta }\left( \bar{X}^{\bar{0}}\right) ^{2}-\chi
\bar{a}_{\bar{\gamma }\bar{\eta }}\bar{X}^{\bar{%
\gamma }}\bar{X}^{\bar{\eta }}\left( X^{0}\right) ^{2} \\
+4\chi ^{2}X^{0}\bar{X}^{\bar{0}}\delta ^{\gamma \bar{\eta }%
}a_{\gamma \theta }\bar{a}_{\bar{\eta }\bar{\rho }}X^{\theta }%
\bar{X}^{\bar{\rho }}
\end{array}
\right] }\,,
\end{eqnarray}
\begin{eqnarray}
G^{\alpha \bar{0}} &=&-\frac{2\chi \delta ^{\alpha \bar{\lambda }}%
\bar{a}_{\bar{\lambda }\bar{\xi }}\bar{X}^{\bar{\xi
}}\left( X^{0}\right) ^{2}\bar{X}^{\bar{0}}}{\left[
\begin{array}{l}
\left( X^{0}\bar{X}^{\bar{0}}\right) ^{2}-\chi a_{\gamma \eta
}X^{\gamma }X^{\eta }\left( \bar{X}^{\bar{0}}\right) ^{2}-\chi
\bar{a}_{\bar{\gamma }\bar{\eta }}\bar{X}^{\bar{%
\gamma }}\bar{X}^{\bar{\eta }}\left( X^{0}\right) ^{2} \\
+4\chi ^{2}X^{0}\bar{X}^{\bar{0}}\delta ^{\gamma \bar{\eta }%
}a_{\gamma \theta }\bar{a}_{\bar{\eta }\bar{\rho }}X^{\theta }%
\bar{X}^{\bar{\rho }}
\end{array}
\right] }\,,\nonumber\\
&&  \notag \\
G^{\alpha \bar{\beta }} &=&\delta ^{\alpha \bar{\beta }}-\frac{%
4\chi ^{2}X^{0}\bar{X}^{\bar{0}}\delta ^{\alpha \bar{\lambda }%
}\delta ^{\sigma \bar{\beta }}a_{\sigma \zeta }\bar{a}_{\bar{%
\lambda }\bar{\xi }}X^{\zeta }\bar{X}^{\bar{\xi }}}{\left[
\begin{array}{l}
\left( X^{0}\bar{X}^{\bar{0}}\right) ^{2}-\chi a_{\gamma \eta
}X^{\gamma }X^{\eta }\left( \bar{X}^{\bar{0}}\right) ^{2}-\chi
\bar{a}_{\bar{\gamma }\bar{\eta }}\bar{X}^{\bar{%
\gamma }}\bar{X}^{\bar{\eta }}\left( X^{0}\right) ^{2} \\
+4\chi ^{2}X^{0}\bar{X}^{\bar{0}}\delta ^{\gamma \bar{\eta }%
}a_{\gamma \theta }\bar{a}_{\bar{\eta }\bar{\rho }}X^{\theta }%
\bar{X}^{\bar{\rho }}
\end{array}
\right] }\,.
\end{eqnarray}
By performing the gauge fixing
\begin{eqnarray}
X^{0} &=&\bar{X}^{\bar{0}}=\sqrt{3}M_{P}\,,\nonumber\\
X^{\alpha } &=&yZ^{\alpha }\left( z\right)\,,\nonumber\\
y &=&\bar{y}=1\,,\nonumber\\
Z^{\alpha } &=&z^{\alpha }\,,
\end{eqnarray}
one then, respectively, obtains
\begin{eqnarray}
G_{0\bar{0}} &=&-1+\frac{\chi }{3M_{P}^{2}}\left( a_{\alpha \beta
}z^{\alpha }z^{\beta }+\bar{a}_{\bar{\alpha }\bar{\beta }}%
\bar{z}^{\bar{\alpha }}\bar{z}^{\bar{\beta }}\right)\,,\nonumber\\
G_{0\bar{\beta }} &=&-\frac{2}{\sqrt{3}}\frac{\chi }{M_{P}}\bar{a}%
_{\bar{\beta }\bar{\gamma }}\bar{z}^{\bar{\gamma }}\,,\nonumber\\
G_{\alpha \bar{0}} &=&-\frac{2}{\sqrt{3}}\frac{\chi }{M_{P}}a_{\alpha
\gamma }z^{\gamma }\,,\nonumber\\
G_{\alpha \bar{\beta }} &=&\delta _{\alpha \bar{\beta }}\,,
\label{1-Umbria}
\end{eqnarray}
\begin{eqnarray}
G^{0\bar{0}} &=&-\frac{1}{\left[ 1-\frac{\chi }{3M_{P}^{2}}\left(
a_{\gamma \eta }z^{\gamma }z^{\eta }+\bar{a}_{\bar{\gamma }%
\bar{\eta }}\bar{z}^{\bar{\gamma }}\bar{z}^{\bar{%
\eta }}\right) +\frac{4}{3}\frac{\chi ^{2}}{M_{P}^{2}}\delta ^{\gamma
\bar{\eta }}a_{\gamma \theta }\bar{a}_{\bar{\eta }\bar{%
\rho }}z^{\theta }\bar{z}^{\bar{\rho }}\right] }\,,\nonumber\\
&&  \notag \\
G^{0\bar{\beta }} &=&-\frac{2\sqrt{3}M_{P}\chi \delta ^{\lambda
\bar{\beta }}a_{\lambda \xi }z^{\xi }}{\left[ 3M_{P}^{2}-\chi \left(
a_{\gamma \eta }z^{\gamma }z^{\eta }+\bar{a}_{\bar{\gamma }%
\bar{\eta }}\bar{z}^{\bar{\gamma }}\bar{z}^{\bar{%
\eta }}\right) +4\chi ^{2}\delta ^{\gamma \bar{\eta }}a_{\gamma \theta }%
\bar{a}_{\bar{\eta }\bar{\rho }}z^{\theta }\bar{z}^{%
\bar{\rho }}\right] }\,,\nonumber\\
&&  \notag \\
G^{\alpha \bar{0}} &=&-\frac{2\sqrt{3}M_{P}\chi \delta ^{\alpha
\bar{\lambda }}\bar{a}_{\bar{\lambda }\bar{\xi }}%
\bar{z}^{\bar{\xi }}}{\left[ 3M_{P}^{2}-\chi \left( a_{\gamma \eta
}z^{\gamma }z^{\eta }+\bar{a}_{\bar{\gamma }\bar{\eta }}%
\bar{z}^{\bar{\gamma }}\bar{z}^{\bar{\eta }}\right)
+4\chi ^{2}\delta ^{\gamma \bar{\eta }}a_{\gamma \theta }\bar{a}_{%
\bar{\eta }\bar{\rho }}z^{\theta }\bar{z}^{\bar{\rho }}%
\right] }\,,\nonumber\\
&&  \notag \\
G^{\alpha \bar{\beta }} &=&\delta ^{\alpha \bar{\beta }}-\frac{%
4\chi ^{2}\delta ^{\alpha \bar{\lambda }}\delta ^{\sigma \bar{%
\beta }}a_{\sigma \zeta }\bar{a}_{\bar{\lambda }\bar{\xi }%
}z^{\zeta }\bar{z}^{\bar{\xi }}}{\left[ 3M_{P}^{2}-\chi \left(
a_{\gamma \eta }z^{\gamma }z^{\eta }+\bar{a}_{\bar{\gamma }%
\bar{\eta }}\bar{z}^{\bar{\gamma }}\bar{z}^{\bar{%
\eta }}\right) +4\chi ^{2}\delta ^{\gamma \bar{\eta }}a_{\gamma \theta }%
\bar{a}_{\bar{\eta }\bar{\rho }}z^{\theta }\bar{z}^{%
\bar{\rho }}\right] }\,.  \label{2-Umbria}
\end{eqnarray}
In particular, the metric $G^{\alpha \bar{\beta }}$ given by Eq.
(\ref{2-Umbria}) is the metric appearing in Eq. (\ref{VJ}). Notice that clearly $%
G^{\alpha \bar{\beta }}$ given by (\ref{2-Umbria}) is not the inverse of
$G_{\alpha \bar{\beta }}$ given by (\ref{1-Umbria}), because what really
holds is
\begin{equation}
G^{0\bar{\beta }}G_{0\bar{\gamma }}+G^{\alpha \bar{\beta }%
}G_{\alpha \bar{\gamma }}=\delta _{\bar{\gamma }}^{\bar{\beta
}}\,.
\end{equation}

\section{The $h$ and $ s$ parts of the NMSSM potential in the Jordan and the Einstein frames }
 \label{app:hands}
Here we present some details of the scalar-gravity part of the supergravity action  given in Eq.  (\ref{action}). We apply it to  the frame function (\ref{Phi1}) and we consider $\beta=\pi/4, \alpha=\alpha_1=\alpha_2=0$.
The kinetic
scalar terms in the Jordan frame are canonical, except for the  contribution to the gauge singlet one due to $\zeta (S\bar S)^2$ corrections to the NMSSM frame function:
\begin{equation}
\mathcal{L}_{J}^{\rm kinetic}=-{\sqrt{-g_{J}}\over 2}\left[ (1- 2\zeta s^2) (\partial
_\mu s)^2+(\partial_\mu h_{1})^2+(\partial_\mu h_2)^2\right] \, ,
\end{equation}
where $ h_{1}\equiv h\cos \beta \, ,  h_2\equiv h\sin \beta $. Note that
the $\zeta$ correction to the kinetic term of the $s$ field is always
small compared to 1 and can be safely neglected. Along the $D$-flat
direction with $\sin (2\beta )=1$  the curvature term in the action for
real fields $h$ and $s$ is
\begin{equation}
\mathcal{L}_{J}^{\rm curv}={\frac{\sqrt{-g_{J}}}2}\left[ 1-{
\textstyle\frac{1}{6}}\left( s^2+h^2\right)
+\ft14\chi h^2 +{
\textstyle\frac{1}{12}} \zeta s^4\right] R(g_{J})\,.
\end{equation}
The potential in the Jordan  frame  for the nonvanishing $\chi$ and $\zeta$ and nonvanishing field $s$ is complicated:
\begin{eqnarray}
V_{J}\left( s,h;\chi ,\lambda ,\rho ,\zeta \right) =  \frac{A_0 + s^2 A_2 + s^4 A_4 + s^6 A_6 + s^8 A_8 }{1-
 2\zeta \,  s^2   + {1\over 3} \zeta \, G \, s^4  }\,.
&&  \label{V_J}
\end{eqnarray}
where we introduced the following notation:
\begin{eqnarray}
&&A_0\equiv  { \lambda
^{2}\over 16}h^{4}, \quad A_2
\equiv  - {h^{2}\over 4} \left ( \left| \lambda \rho \right|  +G \lambda ^{2} ( \chi h^2-4) \right ),  \quad A_4
\equiv {\rho ^{2}\over 4}  -  {\zeta  \,  G h^{2}  \lambda ^{2}\over 8} \left  (32-\left( 12\chi +1/3 \right) h^{2} \right)\,,
\nonumber\\
&&A_6\equiv    {1\over 12} \zeta   \,  G
\left( \lambda ^{2}-\left| \lambda \rho \right| -6\chi \left| \lambda \rho \right| \right ) h^2
\, ,
\quad
A_8
\equiv
{1\over 12} \zeta \, G
\rho ^{2}\, , \quad G\equiv {2\over 8   +\left( 3\chi -2\right) \chi h^{2}} \ .
 \end{eqnarray}
At $\zeta=0$ the potential simplifies to the form given by $ V_J=
G^{\alpha \bar \beta}W_\alpha  \bar W_{\bar \beta} $ when only the $\chi$
term breaks superconformal symmetry of the matter
\begin{eqnarray}
V_{J}\left( s,h;\chi ,\lambda ,\rho ,\zeta=0  \right) =  { \lambda
^{2}\over 4}h^{4}   - {h^{2}\over 4} \left ( \left| \lambda \rho \right|  +2 G \lambda ^{2} ( \chi h^2-2) \right ) s^2  +  {\rho ^{2}\over 4}  s^4 \ .
\end{eqnarray}
At $s=0$ the Jordan potential $
V_J|_{S=0}= G^{S \bar S}W_S  \bar W_{\bar S}
$ {\it restores the superconformal form} at any values of $\chi ,\lambda ,\rho$, and $\zeta$.
\begin{eqnarray}
V_{J}\left( s=0 ,h;\chi ,\lambda ,\rho ,\zeta  \right) =  { \lambda
^{2}\over 4}h^{4}    \ .
\end{eqnarray}
In the Einstein frame for the real fields, the expression of the $F$-term potential
 is
\begin{equation}
V_{E}\left( s,h;\chi ,\lambda ,\rho ,\zeta \right)  =\frac{9}{\Phi^2}  V_{J}= \frac{A_0 + s^2 A_2 + s^4 A_4 + s^6 A_6 + s^8 A_8 }{ \left[ 1 + {1\over 4}\chi  h^{2} -{1\over 6} (s^{2}+ h^{2})  +{1\over 12} \zeta s^{4}\right]
^{2} \, (1-
 2\zeta \,  s^2   + {1\over 3} \zeta \, G \, s^4) },  \label{V_E}
\end{equation}
where all notations are given above.
For $\zeta =0$   the potential in  (\ref{V_E}) reduces to the one studied in  \cite{Ferrara:2010yw}
\begin{equation}
V_{E}\left( s,h;\chi ,\lambda ,\rho ,\zeta=0 \right)  ={\frac{{\frac{\lambda^2}{4}}
h^{4}-|\lambda \rho |s^2h^2-{\frac{2\lambda^2s^2h^2(\chi h^2-4)
}{8+3\chi^2h^2-2\chi h^2 }}+\rho^2|s|^{4}}{4\left[ 1-
\ft16\left( s^2+h^2\right) +\ft14
\chi h^2\right]^2}}\,.
  \label{oldV_E}
\end{equation}
From this equation it is obvious that  at large $\chi h^2$ the mass squared of the $s$ field is negative at $\zeta=0$, as explained in detail in \cite{Ferrara:2010yw}. The crucial positive contribution to the gauge singlet mass term comes from the negative term $-2\zeta s^2$ in the second bracket in the denominator of  (\ref{V_E}), where ${1\over 3} \zeta (S\bar S)^2$ is the quartic correction to the \K\, potential suggested in \cite{Lee:2010hj} and also studied before in \cite{Kitano:2006wz}.

\section{The mass of the charged Higgs field}\label{chargemass}
The mass of the physical
charged Higgs field in globally supersymmetric NMSSM is  \textit{e.g.} given in
\cite{Ellwanger:2009dp} in Eq. (2.29). In the absence of the soft
breaking terms the $F$-term and the $D$-term contribution to mass (in our
notation) is
\be\label{m}
 (m^2_{\pm})_{susy}={1\over 8} ( -\lambda^2 h^2+ {g_2^2\over 2} h^2)\,.
\ee
For $g_2^2>  2\lambda^2$,  the charged Higgs field is not tachyonic, i.e. it is stabilized at   $h_{\pm} = 0$.

In supergravity we  look first at the $s=0$ expression of the
$F$-term potential given by Eq. (\ref{oldV_E}):
\begin{equation}
V_{E}^F \left( s=0,h;\chi ,\lambda  \right)  = \frac{{ \lambda
^{2}\over 16}h^{4}  }{ \left[ 1 + {1\over 4}\chi  h^{2} -{1\over 6}  h^{2}  \right]
^{2}  }\,.  \label{V_E2}
\end{equation}
As explained in Sec. \ref{ss:InflsugraNMSSM}, we can extend this
potential using the $SU(2)$ symmetry to include the charged Higgs, so
that $h^2\rightarrow h^2- h_{\pm}^2$ in the  part of the potential associated
with the holomorphic functions, \textit{i.e.} in the terms originating
from the superpotential and from the $\chi$ terms in the \K\, potential.
However, one has to replace $h^2\rightarrow h^2+ h_{\pm}^2$ in the
$\chi$-independent part of the \K\, potential. This leads to the
following potential
\begin{equation}
V_{E}^F\left( s=0,h; h_{\pm}, \chi ,\lambda  \right)  = \frac{{ \lambda
^{2}\over 16}(h^{2}- h_{\pm}^2)^2  }{ \left[ 1 + {1\over 4}\chi  (h^{2}- h_{\pm}^2) -{1\over 6} (h^{2} + h_{\pm}^2)  \right]
^{2}  }\,.  \label{VF}
\end{equation}
On the other hand, the $D$-term potential depends on charged Higgs field as
follows:
\begin{eqnarray}\label{VD}
V_{E}^{D}\left( h;h_{\pm },g,\chi \right) =\frac{%
g^{2}h^{2}h_{\pm }^{2}}{16\left[ 1+\frac{1}{4}\chi \left( h^{2}-h_{\pm
}^{2}\right) -\frac{1}{6}\left( h^{2}+h_{\pm }^{2}\right) \right] ^{2}}\  .
\end{eqnarray}
One can easily see that \textit{after} inflation, for $\chi h^2\ll 1$, the mass squared of the charged Higgs field $m^2_{\pm}$ coincides with its  value in
the globally supersymmetric case, and therefore the stability condition requires that $g_2^2>  2\lambda^2$.
\textit{During} inflation, one has $h^{2} \ll 1\ll \chi h^2$, and the second
derivatives of (\ref{VF}) and (\ref{VD}) respectively read
\be
V^{''F}_{\pm} \sim - {16 \lambda^2\over \chi^3 h^4} \ , \qquad  V^{''D}_{\pm} \sim  { g^2_2\over \chi^2 h^2} \ .
\ee
In this regime, for $g_2^2>  2\lambda^2$, the $D$-term contribution to $m^2_{\pm}$ is much greater than the $F$-term contribution.
In order to calculate $m^2_{\pm}$, one should also take into account that the kinetic terms for the fields $h_{\pm}$ are not canonical. By doing so, one finds that during inflation
\be
m^2_{\pm} \sim  { g^2_2\over  2 \chi } \gg H^{2} =  {\lambda^2\over  3 \chi^2}\ .
\ee
This means that this field is strongly stabilized at $h_{\pm} = 0$. In other words, under the condition $g_2^2>  2\lambda^2$ the charged Higgs field vanishes \textit{during} and \textit{after} inflation.

\end{document}